\pdfoutput=1
\documentclass{jfm}
\usepackage{graphicx}
\usepackage{epstopdf, epsfig}
\usepackage{natbib}
\usepackage{bm}
\usepackage[dvipsnames]{xcolor}

\shorttitle{Particles clustering and settling in homogeneous turbulence}
\shortauthor{A. J. Petersen, L. Baker and F. Coletti}
\errorstopmode
\title{Experimental study of inertial particles clustering and settling in homogeneous turbulence}

\author{Alec J. Petersen\aff{1}\aff{2}
	\corresp{\email{pet00105@umn.edu}},
	Lucia Baker\aff{1}\aff{2}
	\and Filippo Coletti\aff{1}\aff{2}}

\affiliation{\aff{1}Department of Aerospace Engineering \& Mechanics, University of Minnesota,
	Minneapolis, MN 55414, USA
	\aff{2}St. Anthony Falls Laboratory, University of Minnesota, Minneapolis, MN 55414, USA}

\begin{document}
	\maketitle
\begin{abstract}
	We study experimentally the spatial distribution, settling, and interaction of sub-Kolmogorov inertial particles with homogeneous turbulence. Utilizing a zero-mean-flow air turbulence chamber, we drop size-selected solid particles and study their dynamics with particle imaging and tracking velocimetry at multiple resolutions. The carrier flow is simultaneously measured by particle image velocimetry of suspended tracers, allowing the characterization of the interplay between both the dispersed and continuous phases. The turbulence Reynolds number based on the Taylor microscale ranges from $Re_{\lambda}\approx  200$ - $500$, while the particle Stokes number based on the Kolmogorov scale varies between $St_{\eta} = O(1)$ and $O(10)$. Clustering is confirmed to be most intense for $St_{\eta} \approx 1$, but it extends over larger scales for heavier particles. Individual clusters form a hierarchy of self-similar, fractal-like objects, preferentially aligned with gravity and sizes that can reach the integral scale of the turbulence. Remarkably, the settling velocity of $St_{\eta} \approx 1$ particles can be several times larger than the still-air terminal velocity, and the clusters can fall even faster. This is caused by downward fluid fluctuations preferentially sweeping the particles, and we propose that this mechanism is influenced by both large and small scales of the turbulence. The particle-fluid slip velocities show large variance, and both the instantaneous particle Reynolds number and drag coefficient can greatly differ from their nominal values. Finally, for sufficient loadings, the particles generally augment the small-scale fluid velocity fluctuations, which however may account for a limited fraction of the turbulent kinetic energy. 
\end{abstract}

\section{Introduction}
The dynamics of particles carried by turbulent fluid flows is rich with fascinating phenomena. This is true even for the highly simplified case we consider: homogeneous incompressible turbulence laden with a dilute concentration of spherical, non-Brownian particles smaller than the Kolmogorov scale. The behavior is especially complex when the particles are inertial without being ballistic, i.e. their aerodynamic response time $\tau_p$ is comparable to some relevant temporal scale of the flow. Several important effects observed in this regime are summarized below, making no claim of being exhaustive. For more details the reader is referred to recent reviews from \citet{PoelmaOoms2006}, \citet{BalaEaton2010}, and \citet{GustavssonMehlig2016}. \\
\indent Inertial particles do not distribute homogeneously in turbulent flows, favoring regions of high strain and low vorticity \citep{Maxey87, SquiresEaton1991a}. Such preferential concentration is consistent with early observations of particles accumulating outside the core of large-scale rollers and vortex rings in shear layers and jets \citep{LazeroLasheras89, LongmireEaton1992, IshimaHishidaMaeda93}, and has been attributed to the action of vortices centrifuging the particles outside of their core \citep{EatonFessler94}. While this tendency is noticeable even for weakly inertial particles, it is most intense when the Stokes number based on the Kolmogorov time scale ($St_{\eta} = \tau_p/\tau_{\eta}$) is close to unity \citep{WangMaxey1993}. Such a result is understood as a consequence of $\tau_{\eta}$ being the scale at which the highest vorticity occurs, and indeed particles with $St_{\eta} \approx 1$ were reported to form clusters scaling in Kolmogorov units \citep{KulickFesslerEaton1994, Aliseda2002}. This interpretation also implies that more inertial particles should cluster around eddies of larger turnover times, as was indeed shown by \citet{YoshimotoGoto2007} who reported multi-scale preferential concentration from the dissipative to the inertial range. However, clustering of particles with $St_{\eta}$ significantly larger than unity display different features compared to less inertial ones \citep{BecPRL2007}, and mechanisms other than the centrifuging effect have been proposed \citep{GotoVassilicos2008, BraggCollins2014a, IrelandBraggCollins2016a}.\\
\indent An important effect of the particle inertia is that, as they depart from the fluid streamlines, they may collide. The collision probability is a function of both the local concentration and the relative velocity \citep{SundaramCollins1997,WangWexlerZhou2000} the latter being potentially large as particles approach each other from different flow regions \citep{Falkovich2002, WilkinsonMehlig2005, Bewley2013}. When the particles are liquid droplets, collisions can lead to breakup or coalescence, with important implications for, e.g., atmospheric clouds \citep{Shaw2003} and spray combustion \citep{Jenny2012}. Additionally, in the presence of gravity, the drift of heavy particles crossing fluid trajectories decorrelates their motion from the turbulent fluctuations \citep{Yudine59, Csanady63}, reducing the particle velocity autocorrelation and dispersion coefficient \citep{Reeks77, WellsStock83, ElghobashiTruesdell92, SquiresEaton1991b}.\\
\indent Beside concentration effects, a major consequence of turbulence in particle-laden flows is to alter the rate of gravitational settling. In their seminal work, \citet{WangMaxey1993} confirmed by direct numerical simulations (DNS) the ideas put forward by \cite{Maxey87} using Gaussian flow simulations, and identified a mechanism by which inertial particles with $St_{\eta} = O(1)$ oversample downward regions of the turbulent eddies. This process, referred to as preferential sweeping or fast-tracking, can lead to a significant increase in mean fall speed compared to the expected terminal velocity in quiescent or laminar fluids, $W_0 = \tau_pg$ ($g$ is the gravitational acceleration). Other mechanisms by which turbulence may modify the settling velocity have been proposed, as reviewed by \citet{Nielsen93} and \citet{GGW2012}. These include: vortex trapping, by which relatively light particles are trapped in vortical orbits \citep{Tooby1977}; loitering, due to fast-falling particles spending more time in updrafts than in downdrafts \citep{Nielsen93, KawanisiShiozaki2008}; and non-linear drag increase, which may reduce the traveling speed of particles with significant Reynolds number based on their diameter and slip velocity \citep{Mei1991, Mei1994}; all of these can lead to a decrease in mean fall speed, as opposed to preferential sweeping. Additionally, local hydrodynamic interactions between particles were shown to increase fallspeed of bi-disperse suspensions \citep{WangAyalaGrabowski2007}. While the conditions under which these mechanisms manifest are not well-known, enhanced settling due to preferential sweeping appears to be prevalent for sub-Kolmogorov particles with $St_{\eta} = O(1)$ \citep{YangLei1998,Aliseda2002,YangShy2003,DejoanMonchaux2013,Good2014,IrelandBraggCollins2016b,Rosa2016,Baker2017}. Still, there is no consensus on which turbulence scales are most relevant for this process.\\
\indent Though several works reported settling enhancement by turbulence under similar flow conditions, its extent remains an open question. In numerical simulations, the maximum increase of vertical velocity (which most authors found for $St_{\eta} \approx 1$) varies between studies, from about $0.1W_0$ to $0.9W_0$, or between $0.04u^{\prime}$ and $0.16u^{\prime}$, $u^{\prime}$ being the r.m.s. fluid velocity fluctuation \citep{WangMaxey1993,YangLei1998,DejoanMonchaux2013,Bec2014,Good2014,Baker2017}. Laboratory observations have shown unsatisfactory quantitative agreement with simulations, and between each other. \citet{Aliseda2002} found strong increases in settling velocity of spray droplets in grid turbulence, as high as $1.6W_0$ or $0.26u^{\prime}$ for their most dilute case. Later, \citet{YangShy2003, YangShy2005} reported much weaker settling enhancement for solid particles in zero-mean-flow turbulence facilities. This led \citet{Bosse2006}, comparing their simulations to the results of both groups, to speculate on possible sources of errors in the measurements of \citet{Aliseda2002}. \citet{GGW2012} also found dramatic increases of spray droplet fall speed with turbulence, but this was amplified by mean flow effects \citep{Good2014}. In a subsequent study, \citet{Good2014}, using an extensively tested zero-mean-flow apparatus and high-resolution imaging, found levels of settling enhancement comparable with \citet{Aliseda2002}. But they also showed that point-particle DNS at matching conditions yielding only qualitative agreement with the measurements. In a recent field study, \citet{Nemes2017} measured the fall speed of compact snowflakes by high-speed imaging. They estimated the Stokes number of the observed snowflakes as $St_{\eta} \approx 0.1 - 0.4$, and concluded that the settling velocity in atmospheric turbulence was several times larger than the expected still-air fall speed. \\
\indent An aspect of particle-laden turbulent flows in which our understanding is particularly incomplete is the backreaction of the dispersed phase on the fluid. There is substantial evidence that particles can alter the turbulent fluctuations, but it is still debated under which conditions these will be excited or inhibited \citep{BalaEaton2010}. \citet{GoreCrowe1991} argued that turbulence intensity is augmented/attenuated by particles larger/smaller than one tenth of the integral scale. For fully developed turbulence, this threshold concerns particles significantly larger than the Kolmogorov scale, which will likely modify the turbulence by locally distorting energetic eddies. \citet{Hetsroni1989} proposed a criterion based on the particle Reynolds number $Re_p = d_pU_{slip}/\nu$ (where $d_p$ is the particle diameter, $\nu$ is the kinematic viscosity, and $U_{slip}$ is the slip velocity between both phases), predicting augmented and attenuated turbulence for $Re_p > 400$ and $Re_p < 100$, respectively. These thresholds are also relevant to relatively large particles. \citet{Elghobashi1994} indicated that turbulence modification occurred when the volume fraction $\phi_v$ is higher than approximately $10^{-6}$. In presence of preferential concentration, however, the local volume fraction can be much higher than the mean, enhancing collective effects within and around the clusters \citep{Aliseda2002}. More recently, \citet{Huck2018} showed that by conditioning on the local volume fraction, they could identify three regimes affecting settling velocity: the sparsest dominated by the background flow, the intermediate concentrations suggesting preferential concentration effects, and the densest clusters triggering collective drag. Other parameters have been found to be consequential, including the Stokes number and the particle-to-fluid density ratio $\rho_p/\rho_f$, pointing to the multifaceted nature of the problem \citep{PoelmaOomsWesterweel2007, TanakaEaton2008}. The question of turbulence augmentation versus attenuation is complicated by the fact that the particle-fluid energy transfer is scale-dependent: several studies found that the presence of inertial particles increases the energy at small scales and decreases it at large scales \citep{ElghobashiTruesdell1993, Boivin1998, SundaramCollins1999, Ferrante2003, PoelmaOomsWesterweel2007}. Gravitational settling also contributes to the turbulence modification, as the falling particles transfer their potential energy to the fluid \citep{YangShy2005, HwangEaton2006b, Frankel2016}.\\
\indent Both simulating this class of flows and measuring their properties are challenging. From the computational standpoint, the representation of the particle phase is not straightforward. Since the seminal work of Eaton and Elghobashi, a widespread approach has been to model particles as material points and track their Lagrangian trajectories through the Eulerian flow field obtained by DNS. This led to groundbreaking insights, both when treating the particles as passively advected by the fluid (one-way coupling, Squires \& Eaton 1991a\nocite{SquiresEaton1991a}; Elghobashi \& Truesdell 1992\nocite{ElghobashiTruesdell92}) and when including their backreaction on the fluid (two-way coupling, Squires \& Eaton 1990; Elghobashi \& Truesdell 1993). This approach, however, has well-known limitations. The particle equation of motion assuming Stokes drag (even with the correction terms derived by Maxey \& Riley 1983\nocite{MaxeyRiley83}) is only applicable when particles are much smaller than Kolmogorov scale, and finite-$Re_p$ effects are minimal. Indeed, although the one-way-coupled DNS has led to overall agreement with experiments (e.g., for Lagrangian accelerations, \citet[see][]{Bec2006, Ayyalasomayajula}, quantitative discrepancies indicate that the Stokes drag model may miss important dynamics \citep{Saw2014}. Moreover, modeling the backreaction of the dispersed phase by point-particle methods present technical issues associated with the application of the point-wise forcing on the fluid computational grid \citep{Balachandar2009, Eaton2009, Gualtieri2013}. To overcome these shortcomings, advanced methods have recently been proposed \citep{Gualtieri2015, HorwitzMani2016, IrelandDesjardins2017} whose merits need to be fully appreciated in future comparisons with well-controlled experiments. Setting up the turbulent flow in two-way coupled simulations is also a critical issue: forcing steady-state homogeneous turbulence in either Fourier or physical space leads to artificial energy transfers hardly discernible from the actual interphase dynamics \citep{Lucci2010}. Decaying turbulence leaves the natural coupling unaltered, but quickly drops to low Reynolds numbers and complicates the extraction of statistical quantities.\\
\indent Laboratory measurements also present significant challenges, especially to extract the fluid flow information. Techniques such as Laser Doppler Velocimetry (LDV), Particle Image Velocimetry (PIV) and Particle Tracking Velocimetry (PTV) require tracers that need to be discriminated from the inertial particles. This can be achieved in LDV by signal-processing schemes \citep{RogersEaton1991, KulickFesslerEaton1994}, which are relatively complicated and require careful adjustment \citep{BalaEaton2010}. Moreover, single-point techniques as LDV cannot capture the flow structures and particle clusters that are essential in the dynamics, and as such they have been superseded by whole-field methods, either by conventional or holographic imaging. Time-resolved 3D PTV (often termed Lagrangian Particle Tracking) has been successfully used to investigate inertial particle acceleration \citep{Ayyalasomayajula, Gerashchenko2008, Volk2008}, dispersion \citep{Sabban2011}, relative velocity \citep{Bewley2013, Saw2014}, and collision rates \citep{Bordas2013}. A limitation of this approach is the low particle concentration needed for unambiguous stereo-matching from multiple cameras; this has prevented the volumetric investigation of clustering. For the same reason, experimental studies where both inertial particles and fluid tracers are captured in three dimensions are scarce (\citet{Guala2008} being a rare exception). Two-dimensional (2D) imaging has proven capable of capturing inertial particle distributions and velocities as well as the underlying flow field. Clustering has been probed by 2D imaging since \citet{Fessler1994}, and multiple approaches have since been utilized to characterize concentration fields, as reviewed by \citet{Monchaux2012}. To obtain two-phase measurements, the fluid tracers and inertial particles can be discriminated based on their image size and intensity \citep{KhalitovLongmire2002} and digital/optical filtering \citep{KigerPan2000, PoelmaOomsWesterweel2007}. The motion of the continuous and dispersed phases is then characterized by PIV and PTV, respectively. This approach has allowed the investigation of particle-turbulence interaction in wall-bounded \citep{Paris2001, KigerPan2002, KhalitovLongmire2003} and homogeneous flows \citep{YangShy2003,YangShy2005, HwangEaton2006, PoelmaOomsWesterweel2007, TanakaEaton2010, Sahu2014, Sahu2016}. Many of these studies were obtained for relatively large particles, for which significant loadings are obtained with limited number densities. Sub-Kolmogorov particles are harder to discriminate from tracers, and the velocimetry of the surrounding fluid is especially challenging in presence of clusters (Yang \& Shy 2005). \\
\indent We present the results of an extensive measurement campaign in which sub-Kolmogorov solid particles settle in homogeneous air turbulence created in a zero-mean-flow chamber. Planar imaging at various resolutions is used to probe both dispersed and continuous phases over a wide range of scales, providing insight into several of the outstanding questions discussed above. The paper is organized as follows: the experimental apparatus, the measurement approach, and the parameter space are described in \textsection 2. The particle spatial distribution and clustering are described in \textsection 3, whereas the effect of turbulence on their fall speed is addressed in \textsection 4. In \textsection 5 the simultaneous two-phase measurements are leveraged to explore the particle-turbulence interaction. A discussion of the results and conclusions drawn from them are given in \textsection 6.

\section{Methods}\label{sec:methods}
\subsection{Experimental apparatus}
The experimental facility consists of a chamber where homogeneous air turbulence is produced by jet arrays, and in which known quantities of size-selected solid particles are dropped. The turbulence chamber has been introduced and qualified in \citet{Carter2016} and the unladen flow properties have been investigated in detail in \citet{CarterColetti2017, Carter2018}; here we only give a brief description for completeness. The $5$ m$^3$ closed chamber has acrylic lateral walls and ceiling for optical access, and it contains two facing arrays of 256 quasi-synthetic jets controlled by individual solenoid valves. Using the jet firing sequence proposed by \cite{VarianoCowen2008}, we obtain approximately homogeneous turbulence with negligible mean flow and no shear over a central region of 0.5 by 0.4 by 0.7 m$^3$ in direction $x$, $y$ and $z$, respectively (where $x$ is aligned with the jet axis and $z$ is vertical). This is substantially larger than the integral scale of the flow, allowing for the natural inter-scale energy transfer without major effects of the boundary conditions. Importantly, this also means that the particles, remembering the flow they experience through the history term in their transport equation \citep{MaxeyRiley83}, are not affected by turbulence inhomogeneity. The lack of mean flow (especially small in the vertical direction, \cite[see][]{Carter2016} is beneficial for the unbiased measurement of the particle settling velocity. The properties of the turbulence can be adjusted by varying the average jet firing time $\mu_{on}$, the distance between both jet arrays, and by adding grids in front of the jets. In the present study, we keep the distance at 1.81 m and use a combination of firing times and grids that force turbulence with a Taylor microscale Reynolds number $Re_{\lambda} \approx 200 - 500$; the main properties for the unladen flow are reported in table \ref{tab:flowprops}. These may be altered by the presence of particles, although not dramatically in the considered range of particle types and loadings, as we will discuss in \textsection 5. The flow is anisotropic at all scales, with more intense velocity fluctuations in the $x$ direction \citep{CarterColetti2017}. The fine-scale structure and topology display all signature features of homogeneous turbulence \citep{Carter2018}.\\ 
\begin{table}
	\begin{center}
		\def~{\hphantom{0}}
		\begin{tabular}{cccccccc}
			Grids&$\mu_{on}$ [s] & $u'$ [ms$^{-1}$]&$u'_x/u'_z$ &$L$ [mm]& $\eta$ [mm] &$\tau_{\eta}$ [ms] &$Re_{\lambda}$\\[3pt]
			yes & .4 & 0.34 & 1.41 & 73 & 0.34 & 7.5 & 200\\
			yes & 10.& 0.51 & 1.72 & 100 & 0.28 & 4.9 & 300\\
			no & .2& 0.59 & 1.41 & 90 & 0.27 & 4.8 & 300\\
			no & .4& 0.67 & 1.46 & 99 & 0.24 & 3.8 & 360\\
			no & 3.2& 0.73 & 1.72 & 140 & 0.24 & 3.6 & 500\\
			no &10.& 0.76 & 1.67 & 146 & 0.24 & 3.6 & 500\\
		\end{tabular}
		\caption{Unladen turbulence statistics for the configurations in this study. The r.m.s. velocity $u'$ and the longitudinal integral scale $L$ are based on a weighted average between the $x$ and $z$ directions. The Kolmogorov length scale $\eta = (\nu^3 / \varepsilon)^{1/4}$ and time scale $\tau_{\eta} = (\nu / \varepsilon)^{1/2}$ are based on estimates of the dissipation rate $\varepsilon$ from the 2$^{\mbox{\tiny{\textit{nd}}}}$ order transverse structure functions. For further details \citet[see][]{Carter2016}}
		\label{tab:flowprops}
	\end{center}
\end{table} 
\indent The chamber ceiling is provided with a circular opening (15.2 cm in diameter) connected to a 3 m vertical chute, through which solid particles are introduced at a steady rate using an AccuRate dry material feeder. The feeding rate is adjusted to produce different volume fractions in the chamber. The particles interact with the turbulence for at least 0.7 m before entering the field-of-view. At typical settling rates and depending on the particle types, this corresponds to between tens and hundred of integral time scales of the turbulence-corroborating our observation that the particles spread throughout the chamber quickly upon entering it. We use several types of particles: soda-lime glass beads of various sizes (Mo-Sci Corp.), lycopodium spores (Flinn Scientific, Inc.), and glass bubbles (The 3M Company), all with a high degree of sphericity as verified by optical microscopy. The properties of the considered particle types are listed in table \ref{tab:particle_props}. The aerodynamic response time is iteratively calculated with the Schiller \& Naumann correction \citep{Clift2005}):
\begin{equation} \label{tau_p}
\tau_p = \frac{\rho_pd_p^2}{18\mu (1+0.15Re_{p,0}^{0.687})}
\end{equation}
where $\mu$ is the air dynamic viscosity, $\rho_p$ and $d_p$ are the particle density and mean diameter, and $Re_{p,0} = d_pW_0/\nu$ is the particle Reynolds number based on the still-air settling velocity. \\

\begin{table}
	\begin{center}
		\def~{\hphantom{0}}
		\begin{tabular}{ccccc}
			$d_p $[$\mu$m] & material & $\rho_p$ [kg/cm$^3$] & $\tau_p$ [ms] & $Re_{p,0}$ \\[3pt]
			$91 \pm 11$ & glass bubbles & 100 & $1.7$ & 0.10 \\
			$30 \pm 2$ & lycopodium & 1200 & $3.1$ & 0.06 \\
			$32 \pm 7$ & glass & 2500 & $7.4$ & 0.15 \\
			$52 \pm 6.1$ & glass & 2500 & $17$ & 0.56 \\
			$96 \pm 11$ & glass & 2500 & $47 $ & 3.26 \\
		\end{tabular}
		\caption{Particle properties. The diameters $d_p$ are listed in mean $\pm$ standard deviation.}
		\label{tab:particle_props}
	\end{center}
\end{table}
\subsection{Measurement techniques}
\begin{center}
	\begin{figure}
		\centerline{\includegraphics[width=\textwidth]{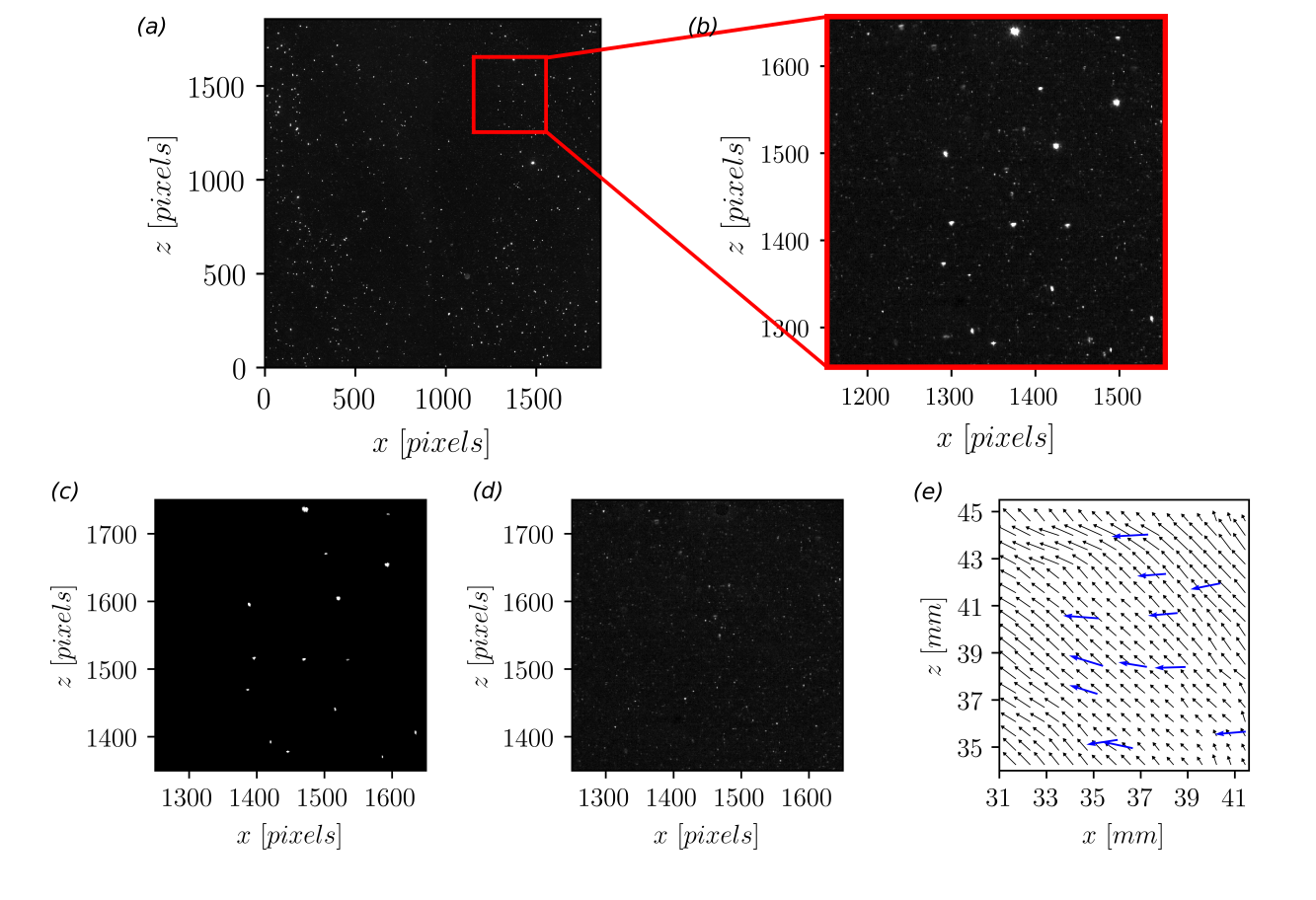}}
		\caption{Image processing procedure for PIV/PTV measurements of tracer and particle motion: (a) raw image, (b) inset of raw image,  (c) particle-only image, (d) tracer-only image, (d) resulting fluctuating fluid and particle velocity vectors.}
		\label{fig:processing}
	\end{figure}
\end{center}

\begin{center}
	\begin{figure}
		\centerline{\includegraphics[width=\textwidth]{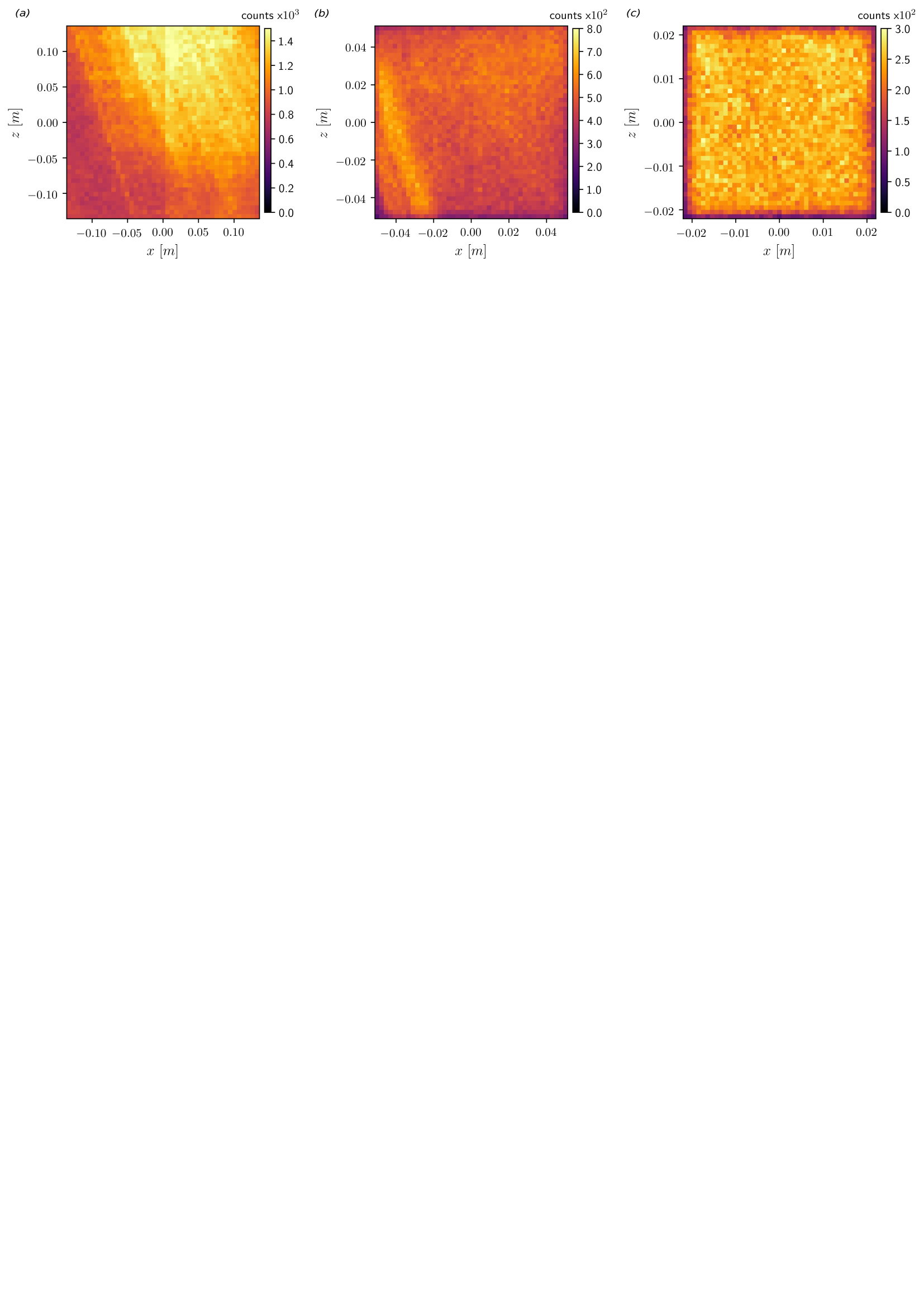}}
		\caption{Particle counts for example fields of view.}
		\label{fig:concentration_maps}
	\end{figure}
\end{center}

All measurements are performed along the $x-z$ symmetry plane, using the same hardware as in \citet{Carter2016} et al. (2016) and \citet{CarterColetti2017, Carter2018}. The air flow is seeded with $1-2$ $\mu$m DEHS (di-ethyl-hexyl-sebacate) droplets, which are small enough to faithfully trace the air motion without altering the particle transport. The imaging system consists of a dual-head Nd:YAG laser (532 nm wavelength, 200 mJ/pulse) synchronized with a 4 Megapixel, 12-bit CCD camera. To capture the wide range of spatial scales, we perform measurements using Nikon lenses with focal lengths of 50, 105, and 200 mm, yielding a range of fields of view (FOV) and resolutions reported in table 3. The laser pulse separation, chosen as a compromise to capture the in-plane motion of both flow tracers and inertial particles, ranges between $110 \mu$s (to image the heavier particles in the smaller FOV and stronger turbulence) and $850$ $\mu$s (for the lighter particles in the larger FOV in weaker turbulence). For all measurements the typical displacement of both tracers and inertial particles is approximately 5 pixels. For the inertial particles this corresponds to 1 visual particle diameter on average. The sampling frequency is 7.25 Hz, which provides approximately uncorrelated realizations (given the typical large-eddy turnover time between 0.1 and 0.2 s). For most experiments, 2000 image pairs are recorded. For the cases with highest loading, the finite supply of particles in the screw-feeder limits the recordings to 1000 -- 1500 image pairs. Because those cases also have the highest number of particles per image (and do not allow accurate fluid measurements), the statistical convergence of the reported quantities is not significantly altered.\\
\indent The two-phase flow images are used to perform simultaneous PIV on the tracers and PTV on the inertial particles. After subtracting a pixel-wise minimum intensity background from each image, both phases are separated via an algorithm inspired by \citet{KhalitovLongmire2002}. We set all pixels below a threshold intensity to zero, which we choose based on a visual inspection of each individual case. Contiguous groups of non-zero pixels are identified and labeled as either tracers or inertial particles contingent on their position in a size-intensity map. Pixels belonging to inertial particles are subtracted and substituted with a Gaussian noise of the same mean and standard deviation as the corresponding image. The resulting tracer-only images are then processed via a cross-correlation PIV algorithm with iterative window offset and deformation, applying one refinement step and 50\% overlap \citep{Nemes2015}. A Gaussian fitting function is used to determine sub-pixel displacements. Tests on synthetic images confirm that the cross-correlation algorithm accurately predicts tracer displacements within $\pm$0.1 pixels when their images are 2--3 pixels in size, representative of our small-FOV recordings. For the large and intermediate FOV, moderate peak-locking is present in the distribution of particle displacements. This only marginally affects the fluid statistics in the present zero-mean-flow configuration, since the entire range of pixel displacement is associated with the turbulent motion \citep{Carter2016}. PIV vector validation is based on signal-to-noise ratio and deviation from the median of the neighboring vectors \citep{WesterweelScarano2005}. Non-valid vector percentages vary between experiments depending on camera resolution and inertial particle volume fraction. While most runs have fewer than 8\% of vectors rejected, the cases with higher particle concentration have 15--18\% rejected fluid velocity vectors. This is due to the background noise from light scattered by the ensemble of the inertial particles, rather than to the removal of individual particle images from local interrogation windows. Indeed for the present cases, no statistical correlation is found between the location of non-valid vectors and the inertial particle position, except for the most highly concentrated cases where we do not attempt to extract fluid information. The dominant source of uncertainty on the flow statistics is the finite sample size, yielding typical uncertainties of 3\% for mean velocity measurements and 5\% for root mean square (r.m.s.) velocity fluctuations \citep{BendatPiersol2011}l.\\
\begin{table}
	\begin{center}
		\def~{\hphantom{0}}
		\begin{tabular*}{\textwidth}{c @{\extracolsep{\fill}} ccccc}
			Focal & Field of view & Field of view & Resolution  & PIV vector spacing & PIV vector spacing \\
			length (mm) & (cm$^2$) & ($L_{L,1}^2$) & (pix/mm) & (mm) &($\eta$)\\[3pt] \hline
			50 & $30^2$ & $2^2-3.3^2$ & 6 & N/A & N/A \\
			105 & $12.5^2$ & $0.8^2 - 1.6^2$ &15 & $0.88-1.77$&$3.3-7.4$\\
			200 & $4.5^2$ & $0.3^2-0.6^2$ &40& $0.33-0.61$ & $1.1-2.6$\\
		\end{tabular*}
		\caption{Imaging parameters obtained with the CCD camera (2048 by 2048 pixels, 7.4 $\mu$m pixel size) when mounting the different lenses. The large FOV with the 50 mm lens is not used for PIV.}
		\label{tab:FOVs}
	\end{center}
\end{table}
\indent The objects labeled as inertial particles are moved to a blank image and tracked via an in-house PTV algorithm. This is based on the cross-correlation approach \citep{Ohmi2000, Hassan1992}, although our version uses the full 12-bit pixel intensity information rather than the binarized image. The algorithm searches for a matching object within a specified radius around each particle centroid, maximizing the correlation coefficient between the image pairs. It performs well even with multiple neighboring particles, since the local distribution pattern remains similar in the image pair. Mild peak-locking is present in the large-FOV recordings of the smaller particles; however, as in the PIV of the tracers, the zero-mean-flow configuration limits the impact on the measured statistics. The process of phase separation is illustrated in figure \ref{fig:processing}, where a sample image is shown along with the resulting fluctuating velocity of flow tracers and inertial particles from PIV and PTV, respectively.
\indent\indent An advantage of the cross-correlation PTV approach is that its accuracy is weakly affected by the uncertainty in locating the object centroid. The latter will, however, affect the measurement of the particle spatial distribution, which is of interest in our study. We use different methods for locating the centroid, depending on the imaging conditions. For the larger FOV, the inertial particles cover typically 3 -- 4 non-saturated pixels and a standard three-point Gaussian fit is appropriate. For the intermediate and small FOV, the particle images are larger and sometimes saturated. In these cases, we use a least-squares Gaussian fit: for each object, a circular particle image of equivalent size is generated, following a Gaussian spatial distribution centered on the center-of-mass of the original object (the weights being the pixel intensities). The position of the circular particle is then fitted to the original particle through a two-dimensional least-squares regression, yielding the sub-pixel centroid. The accuracy of both the three-point and least-squares methods has been tested on synthetic particle images with and without saturation. The least-squares method is more computationally expensive but more accurate, with an average error of 0.12 pixels in locating the centroid of saturated particles, against 0.45 pixels for the three-point fit. Importantly, the spatial distribution of particle count presents significant inhomogeneities only over the largest FOV, as shown in figure \ref{fig:concentration_maps} for a representative case, which is due to the somewhat uneven laser illumination at those scales. This allows us to compute particle statistics via space-time ensemble-averages over the full window, without the need of compensating for spatial gradients \citep{Sumbekova2017}.\\
\indent Using our particle identification methods, we also are able to estimate the volume fraction $\phi_v$, by counting the number of particles in the field of view and comparing their total volume with the illuminated volume. Even at the higher loading considered, the average inter-particle distance is at least $\sim$1 mm, which is larger than the particle image. Due to clustering, particles may be found much closer to each other, preventing their individual identification. However, intense clustering usually pertains to a limited fraction of the particle set \citep{Baker2017}, and thus the number of undetected ones is expected to be relatively small. \citet{YangShy2005} and recently \citet{Sahu2014, Sahu2016} carried out experiments in similar conditions and used the same approach to estimate $\phi_v$. This method has proven robust also in our recent study of a particle-laden channel flow, in which we imaged 50 $\mu$m glass beads at $\phi_v$ = $O(10^{-5})$ with a similar PIV system \citep{Coletti2016,Nemes2016}. In that case the imaging-based volume fraction agreed to within 12--15\% the value obtained from the amount of particles accumulated in the exit plenum during a given run time.
\subsection{Vorono\"i tessellation and cluster identification}
\begin{center}
	\begin{figure}
		\centerline{\includegraphics[width=\textwidth]{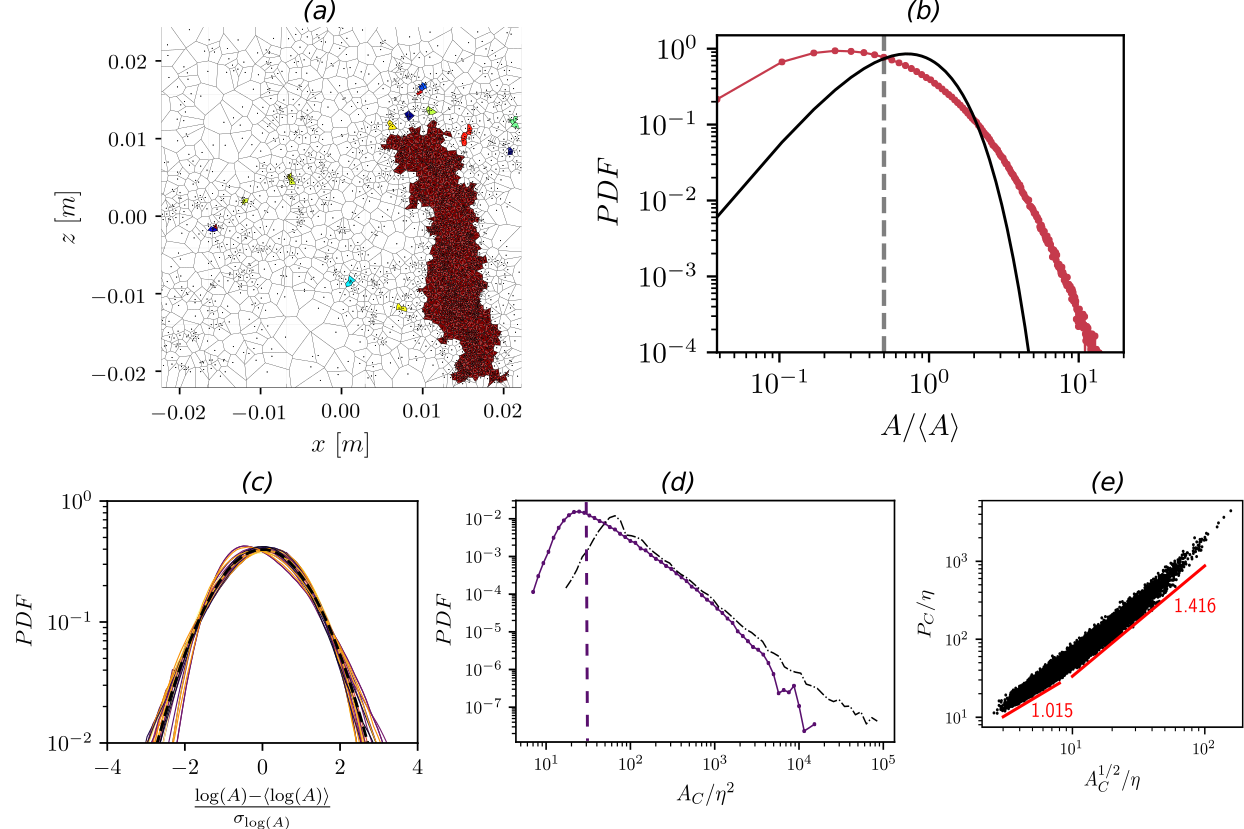}}
		\caption{(a) Example Vorono\"i tessellation with connected sets of Vorono\"i cells below the area threshold colored. (b) Example Vorono\"i area PDF shows clear departure from random Poisson distribution indicating the presence of clusters and voids for a case with $St_{\eta} = 3.2$. (c) All PDFs of log($A$), centered at the mean and normalized by the standard deviation. The black dotted line shows a normal distribution with variance = 1. (d) PDF of a sample case, with (purple circles) and without (black dashes) the neighboring cell condition. (e) Scatter plot of cluster perimeter and square root of its area (same case as in (d)).}
		\label{fig:voronoi}
	\end{figure}
\end{center}

To analyze the spatial distribution and concentration of the inertial particles, and in particular the properties of discrete clusters, we make use of the Voronoï diagram method \citep{Monchaux2010}. This approach divides the domain (in this case, the two-dimensional image) into a tessellation of cells associated to individual particles, each cell containing the set of points closer to that particle than to any other. The inverse of the area $A$ of each cell equals the local particle concentration, $C = 1/A$. The method has been used to analyze particle-laden turbulent flows in both experimental \citep{Obligado2014, Rabencov2015, Sumbekova2017} and numerical studies \citep{Tagawa2012, Kidanemariam2013, DejoanMonchaux2013, Zamansky2016, Frankel2016, Baker2017,DejoanMonchaux2017}. Figure \ref{fig:voronoi}a shows the Vorono\"i tessellation for one small-FOV realization, and a representative probability density function (PDF) of cell areas normalized by the mean value $\langle A \rangle $ is plotted in figure \ref{fig:voronoi}b. (Here and in the following, angle brackets denote ensemble-average.) As typical for clustered particle fields, the observed PDF is much wider than that of a random Poisson process, which follows a $\Gamma$ distribution \citep{Ferenc2007}. Figure \ref{fig:voronoi}c shows the centered and normalized PDFs of the logarithm of the Vorono\"i cell areas for all our experiments, indicating a reasonable collapse that emphasizes their quasi-lognormality. This behavior has been exploited to characterize the particle distribution by a single parameter, the standard deviation $\sigma_A$ \citep{Monchaux2010}.\\
\indent The value $A^*$, below which the probability of finding sub-average cell areas is higher than in a Poisson process, is usually taken as the threshold for particles to be considered clustered (Monchaux \textit{et al.} 2010, Rabencov \& van Hout 2015, Sumbekova \textit{et al.} 2017, among others). Individual clusters are then defined as connected groups of such particles, as shown in figure \ref{fig:voronoi}a. To avoid spurious edge effects, we apply the additional constraint that the area of all neighboring cells are also smaller than $A^*$ (a condition first introduced by \citet{Zamansky2016}. Figure \ref{fig:voronoi}d shows the PDFs of cluster areas $A_C$ for a representative case, obtained with and without this latter condition. Such a condition separates objects connected by only one Voronoi cell. The edge effect produces ripples in the distribution without the neighbor cell condition, indicating that certain cluster sizes are unlikely to occur, possibly due to the coagulation of neighboring connected objects; the application of the neighboring cell condition removes this artifact.  This allows us to isolate very small clusters and to separate artificially large ones; hence the apparent shift in the area PDF.\\
\indent Following \citet{Baker2017}, we use the Vorono\"i diagram method to identify individual clusters, focusing on those sufficiently large to exhibit a scale-invariant structure. Figure \ref{fig:voronoi}e displays, for the same case as in figure \ref{fig:voronoi}d, the scatter plot of cluster perimeters ($P_C$) versus the square root of their areas ($A_C^{1/2}$). (We refer to ‘cluster perimeter’ and ‘cluster area’, although these are strictly properties of the connected set of Vorono\"i cells associated to the particles in each cluster, rather than to the cluster itself). For small clusters, the data points follow a power law with exponent $\sim$1 as expected for regular two-dimensional objects, while for larger ones the exponent is approximately 1.4, indicating a convoluted structure of the cluster borders. This trend, common to all our experiments, was observed in several previous studies \citep[e.g.,][]{Monchaux2010, Rabencov2015, Baker2017} using Vorono\"i tessellation, and earlier \citet{Aliseda2002} using box-counting), and is consistent with the view of inertial particle clusters as fractal sets \citep{Bec2003, BecPRL2007, Calzavarini2008}, although it must be remarked that this latter feature was shown to be associated to the dissipative scales. The minimum size for the emergence of fractal clustering is difficult to identify precisely in figure \ref{fig:voronoi}e; however, as shown by \citet{Baker2017}, this also corresponds to the emergence of self-similarity of the cluster sizes, as indicated by the power-law decay in their size distribution. This threshold can be located with more confidence in figure \ref{fig:voronoi}d (dashed line); it is taken as the condition for a cluster to be “coherent”, i.e. associated to the coherent motions in the underlying turbulent field rather than by accidental particle proximity \citep{Baker2017}.

\subsection{Parameter space}
Table 4 reports the main physical parameters and imaging resolution for all experimental runs, 57 in total. Not all cases are used for all types of analysis: for example, the glass bubbles are very light and do not allow sufficiently accurate measurements of the settling velocity, while the 100 $\mu$m glass beads do not disperse homogeneously enough to perform clustering analysis. The importance of particle weight is characterized by the settling parameter $Sv_{\eta} = W_0/u_{\eta}$, where $u_{\eta} = \eta/\tau_{\eta}$ is the Kolmogorov velocity. Since both small-scale and large-scale eddies are consequential for the settling process, a definition based on the r.m.s. fluid velocity fluctuation, $Sv_L = W_0/u^{\prime}$, is also relevant \citep{Good2014}. The Froude number $Fr = St_{\eta}/Sv_{\eta}$ is also often used in literature, and is reported in table 4 for completeness. From a comparison with previous studies, we expect the turbulence to induce significant clustering and settling modification, and the particles to possibly modify the turbulence at the higher volume fractions. We remark that, as in any laboratory study with a fixed gravitational acceleration, varying only one parameter at a time is not feasible. For example, increasing $St_{\eta}$ by using heavier particles leads to higher $Sv_{\eta}$, unless $Re_{\lambda}$ is also adjusted. Likewise, varying $\phi_v$ may modify the turbulence properties, and therefore the effective values of the other parameters. Therefore, throughout the paper we will often show the simultaneous dependence of the observables with multiple parameters.\\
\begin{table}
	\begin{center}
		\def~{\hphantom{0}}
		\begin{tabular*}{\textwidth}{c @{\extracolsep{\fill}} lcccccccr}
			Case &material &$Re_{\lambda}$ &$d_p/\eta$&$St_{\eta}$ &	$Sv_{\eta}$& $Sv_L$& $Fr$ & $\phi_v$&$\phi_m$\\ \hline
1 &	glass& 	500& 	0.52& 	20.8& 	5.6& 	0.59& 	3.7& 	5.5e-5& 	1.0e-1\\
2 &	glass& 	500& 	0.43& 	14.7& 	6.5& 	0.66& 	2.3& 	2.6e-6& 	5.3e-3\\
3 &	glass& 	300& 	0.43& 	14.0& 	6.8& 	0.77& 	2.1& 	3.4e-5& 	6.9e-2\\
4 &	glass& 	500& 	0.42& 	14.0& 	6.7& 	0.66& 	2.1& 	2.5e-6& 	5.2e-3\\
5 &	glass& 	500& 	0.40& 	12.4& 	7.1& 	0.63& 	1.8& 	1.5e-5& 	3.1e-2\\
6 &	glass& 	500& 	0.40& 	12.4& 	7.1& 	0.63& 	1.8& 	1.6e-5& 	3.7e-2\\
7 &	glass& 	300& 	0.40& 	12.5 &	7.1& 	0.90& 	1.8& 	1.6e-6& 	3.4e-3\\
8 &	glass& 	300& 	0.39& 	12.0&	7.2& 	0.86& 	1.7& 	1.7e-6& 	3.8e-3\\
9 &	glass& 	300& 	0.35& 	9.8& 	8.0& 	0.77& 	1.2& 	1.7e-5& 	3.5e-2\\
10 &	glass& 	500& 	0.26& 	6.5& 	2.3& 	0.24& 	2.9& 	6.5e-6& 	1.3e-2\\
11 &	glass& 	500& 	0.26& 	6.6& 	2.3& 	0.23& 	2.7& 	3.2e-7& 	6.5e-4\\
12& 	glass& 	400& 	0.26& 	6.4& 	2.3& 	0.26& 	2.8& 	2.4e-7& 	4.9e-4\\
13 &	glass& 	500& 	0.24& 	5.8& 	2.3& 	0.23& 	2.5& 	1.2e-6& 	2.4e-3\\
14 &glass& 	500& 	0.24& 	5.4& 	2.4& 	0.24& 	2.2& 	9.5e-7& 	1.9e-3\\
15& 	glass& 	500& 	0.23& 	5.2& 	2.5& 	0.23& 	2.1& 	1.4e-6& 	2.8e-3\\
16& 	glass& 	500& 	0.23& 	5.1& 	2.6& 	0.25& 	2.0& 	1.4e-7& 	2.9e-4\\
17& 	glass& 	300& 	0.23& 	5.1& 	2.5& 	0.30& 	2.0& 	3.6e-7& 	7.4e-4\\
18& 	glass& 	300& 	0.23& 	4.9& 	2.6& 	0.31& 	1.9& 	2.2e-5& 4.6e-2\\
19& 	glass& 	500& 	0.22& 	4.6& 	2.6& 	0.23& 	1.8& 	2.3e-6& 4.8e-3\\
20& 	glass& 	500& 	0.22& 	4.6& 	2.6& 	0.23& 	1.8& 	6.6e-7& 	1.3e-3\\
21& 	glass& 	500& 	0.22& 	4.6& 	2.6& 	0.23& 	1.8& 	2.8e-5& 	5.7e-2\\
22& 	glass& 	500& 	0.21& 	4.4& 	2.7& 	0.26& 	1.7& 	3.1e-6& 	6.3e-3\\
23& 	glass& 	300& 	0.21& 	4.2& 	2.7& 	0.34& 	1.5& 	2.3e-6& 	4.8e-3\\
24& 	glass& 	300& 	0.21& 	4.6& 	2.6& 	0.28& 	1.7& 	1.8e-6& 	3.8e-3\\
25& 	glass& 	500& 	0.20& 	4.1& 	2.7& 	0.30& 	1.5& 	6.6e-8& 	1.4e-4\\
26& 	glass& 	300& 	0.20& 	4.1& 	2.8& 	0.34& 	1.5& 	1.0e-6& 	2.0e-3\\
27& 	glass& 	300& 	0.19& 	3.6& 	2.9& 	0.29& 	1.2& 	6.5e-6& 	1.3e-2\\
28& 	glass& 	300& 	0.19& 	3.6& 	2.9& 	0.29& 	1.2& 	3.8e-6& 	7.8e-3\\
29& 	glass& 	300& 	0.19& 	3.2& 	3.3& 	0.38& 	1.0& 	2.3e-7& 	4.7e-4\\
30& 	glass& 	300& 	0.18& 	3.3& 	3.0& 	0.39& 	1.0& 	1.6e-7& 	3.2e-4\\
31& 	glass& 	500& 	0.15& 	2.8& 	1.0& 	0.10& 	2.9& 	4.8e-7& 	9.8e-4\\
32& 	glass& 	500& 	0.15& 	3.2& 	0.90& 	0.10& 	3.5& 	3.0e-7& 	6.2e-4\\
33& 	glass& 	500& 	0.15& 	2.9& 	1.0& 	0.10& 	3.0& 	4.4e-8& 	9.0e-5\\
34& 	glass& 	300& 	0.14& 	2.4& 	1.1& 	0.13& 	2.2& 	8.0e-8& 	1.6e-4\\
35& 	glass& 	500& 	0.13& 	2.4& 	1.0& 	0.10& 	2.3& 	8.0e-8& 	1.6e-4\\
36& 	glass& 	300& 	0.13& 	2.2& 	1.1& 	0.13& 	2.0& 	4.2e-7& 	8.5e-4\\
37& 	glass& 	500& 	0.12& 	2.0& 	1.1& 	0.10& 	1.8& 	5.8e-7& 	1.2e-3\\
38& 	glass& 	500& 	0.12& 	2.0& 	1.1& 	0.10& 	1.8& 	2.7e-8& 	5.0e-5\\
39& 	glass& 	500& 	0.12& 	2.0& 	1.1& 	0.10& 	1.8& 	2.6e-6& 	5.3e-3\\
40& 	glass& 	500& 	0.12& 	2.0& 	1.1& 	0.10& 	1.8& 	1.2e-7& 	2.4e-4\\
41& 	glass& 	300& 	0.12& 	2.0& 	1.1& 	0.13& 	1.7& 	6.3e-7& 	1.3e-3\\
42& 	glass& 	300& 	0.12& 	1.8& 	1.2& 	0.14& 	1.6& 	8.4e-7& 	1.7e-3\\
43& 	glass& 	300& 	0.11& 	1.6& 	1.3& 	0.12& 	1.2& 	7.8e-7& 	1.6e-3\\
44& 	glass& 	300& 	0.11& 	1.7& 	1.2& 	0.14& 	1.4& 	8.8e-7& 	1.8e-3\\
45& 	glass& 	300& 	0.11& 	1.7& 	1.2& 	0.15& 	1.4& 	1.4e-7& 	2.8e-4\\
46& 	glass& 	300& 	0.11& 	1.6& 	1.3& 	0.12& 	1.2& 	2.1e-6& 	4.4e-3\\
47& 	glass& 	300& 	0.11& 	1.8& 	1.2& 	0.14& 	1.5& 	6.5e-8& 	1.3e-4\\
48& 	glass& 	300& 	0.11& 	1.6& 	1.3& 	0.13& 	1.2& 	6.7e-8& 	1.4e-4\\
49& 	glass& 	300& 	0.11& 	1.7& 	1.2& 	0.14& 	1.4& 	1.3e-7& 	2.7e-4\\
50& 	glass& 	200& 	0.09& 	1.1& 	1.6& 	0.22& 	0.73& 	1.2e-7& 	2.4e-4\\
51& 	lycopodium& 	500& 	0.12& 	0.80& 	0.46& 	0.04& 	1.8& 	5.4e-7& 	5.3e-4\\
52& 	lycopodium& 	500& 	0.12& 	0.80& 	0.46& 	0.04& 	1.8& 	8.6e-8& 	8.0e-5\\
53& 	lycopodium& 	300& 	0.12& 	0.75& 	0.48& 	0.06& 	1.6& 	1.6e-7& 	5.0e-4\\
54& 	lycopodium& 	300& 	0.11& 	0.63& 	0.51& 	0.05& 	1.2& 	9.4e-6& 	9.2e-3\\
55& 	lycopodium& 	300& 	0.11& 	0.63& 	0.51& 	0.05& 	1.2& 	6.0e-7& 	5.9e-4\\
56& 	lycopodium& 	300& 	0.11& 	0.63& 	0.51& 	0.05& 	1.2& 	1.9e-7& 	1.8e-4\\
57& 	glass bubbles& 	300& 	0.34& 	0.37& 	0.30& 	0.03& 	1.2& 	3.4e-5& 	2.7e-3\\
		\end{tabular*}
		\caption{Main experimental parameters for all considered cases, ordered in decreasing particle Stokers number $St_{\eta}$.}
		\label{tab:params}
	\end{center}
\end{table}
\\
\section{Particle spatial distribution}
\subsection{Particle fields}
In this section we explore the spatial structure of the inertial particle fields and the length scales over which clustering occurs. In the literature this has been characterized by two main tools: the radial distribution function ($RDF$) and, more recently, Vorono\"i tessellation. Both methods provide different and complementary information---we apply them both for a comprehensive description.\\
\indent The RDF describes the scale-by-scale concentration in the space surrounding a generic particle, compared to a uniform distribution. For 2D distributions such as those obtained by planar imaging, this is defined as:
\begin{equation}
g(r) = \frac{N_r / A_r}{N / A_{tot}}
\end{equation}
where $N_r$ represents the number of particles within an annulus of area $A_r$, while $N$ is the total number of particles within the planar domain of area $A_{tot}$. In presence of clustering, the RDF is expected to increase for decreasing $r$, and the range over which it remains greater than unity indicates the length scale over which clustering occurs \citep[e.g., ][]{SundaramCollins1997, ReadeCollins2000, Wood2005, Saw2008, deJong2010, IrelandBraggCollins2016a, IrelandBraggCollins2016b}. We compute RDFs by binning particle pairs based on their separation distance. To avoid projection biases at separations below the illuminated volume thickness \citep{HoltzerCollins2002}, we only calculate $g(r)$ for $r > 1.5$ mm. As noted by \citet{deJong2010}, imaging-based RDF measurements are sensitive to the size and shape of the observation region, and some sort of edge-correction strategy is needed for particles near the image boundaries. One can omit statistics for radial annuli that cross the image boundary, but this approach has two shortcomings: the maximum separation becomes limited to the radius of the domain-inscribed circle; and the number of particle pairs per unit area used to calculate the RDF decreases as the separation increases. Both effects combine to thwart the reliable assessment of large-scale clustering. Indeed, past RDF measurements at distances $O(L)$ in flows with wide scale separation were obtained using single-point probes and invoking Taylor’s hypothesis \citep{Saw2008,Saw2012,BatesonAliseda2012}. Here, following \citet{deJong2010}, we leverage the spatial homogeneity of our fields and apply a periodic-domain correction: the particle field is mirrored across the image boundaries, so that the same number of radial annuli can be used for each particle location, yielding a maximum separation equal to the full size of the FOV. Although this assumption introduces unphysical correlations between particles near the reflected boundaries, numerical experiments using DNS showed the associated error to be small \citep{Salazar2008}.\\
\indent Due to the large range of scales separation ($L/\eta \sim O(10^3)$, $L/d_p \sim O(10^4 - 10^5)$) it is not feasible to simultaneously resolve all scales at play, and thus the small-FOV and large-FOV imaging will suffer from large-scale and small-scale cutoffs, respectively. Comparing the different FOVs, however, provides quantitative information over a wide range of scales. Figure 5a shows examples from the small-FOV measurements. At small separations, several authors have found satisfactory fit to the data using a power law, which indicates a self-similar spatial distribution \citep{Chun2005, Salazar2008, Zaichik2009, IrelandBraggCollins2016a,IrelandBraggCollins2016b}:
\begin{equation}
g(r/\eta) = c_0(r/\eta)^{-c_1}
\end{equation}
where $c_0$ and $c_1$ are coefficients dependent on $St_{\eta}$ (and, in presence of gravity, $Sv_{\eta}$). While theoretical arguments consistent with this formulation strictly apply for dissipative separations ($r/\eta < 1$,\citet{Chun2005}, \citet{Saw2008} argued that the power-law form should continue into the correlation scale of the velocity gradients ($r/\eta = O(10)$). In figure 5a we see indeed that RDFs closely follow a power-law decay up to $r/\eta \approx 40$ for $St_{\eta}$ close to unity. The departure from the power law at larger separations indicate the particle set is not self-similar at those scales \citep{Bragg2015}. We evaluate the coefficients $c_0$ and $c_1$ by least-square fit over the range $10 < r/\eta < 30$, and plot them in figure 5b and 5c as a function of $St_{\eta}$. The error bars for the coefficients comes from the covariance matrix of the fit. The results, which are only weakly sensitive to varying the fit upper bound between $r/\eta = 20$ and $40$, display the higher values in the approximate range $1.5 < St_{\eta} < 4$. This confirms that particles with Stokes number $O(1)$ display the stronger degree of clustering over the near-dissipative range. However, the most intense clustering occurs for $St_{\eta} > 1$, possibly because of the significant effect of gravitational settling as discussed below. The trend and values are in fair quantitative agreement with the DNS of \citet{IrelandBraggCollins2016b} in similar conditions. The $c_1$ coefficient is related to the correlation dimension used in dynamical system theory \citep{Bec2008}, $D_2 = n-c_1$, where the number of spatial dimensions is $n = 2$ for our planar realizations. Figure 5d plots $D_2$ for the different cases, showing trends and values consistent with the channel flow experiments by \citet{Fessler1994} and the grid turbulence experiments by \citet{Monchaux2010, Monchaux2012}. For increasing particle inertia, one expects a loss of spatial correlation as the particle response time grows beyond the fine turbulent scales. Although the present range does not extend to very large $St_{\eta}$, we note that the return to a homogeneous distribution appears slow. Recent numerical studies compared settling and non-settling conditions, and concluded that gravity hinders clustering for $St_{\eta} < 1$ but enhances it for $St_{\eta} > 1$, resulting in significant clustering over a wide range of Stokes numbers \citep{BecHomannRay2014, Gustavsson2014, IrelandBraggCollins2016b, Matsuda2017, Baker2017}. \citet{IrelandBraggCollins2016b} attributed this behavior to the competing effects of the particle path history and preferential flow sampling. \citet{Sahu2016} measured RDFs for spray droplets and also noticed an increasing tendency to cluster for increasing $St_{\eta}$ (although their range was very close to unity).
\begin{center}
	\begin{figure}
		\centerline{\includegraphics[width=\textwidth]{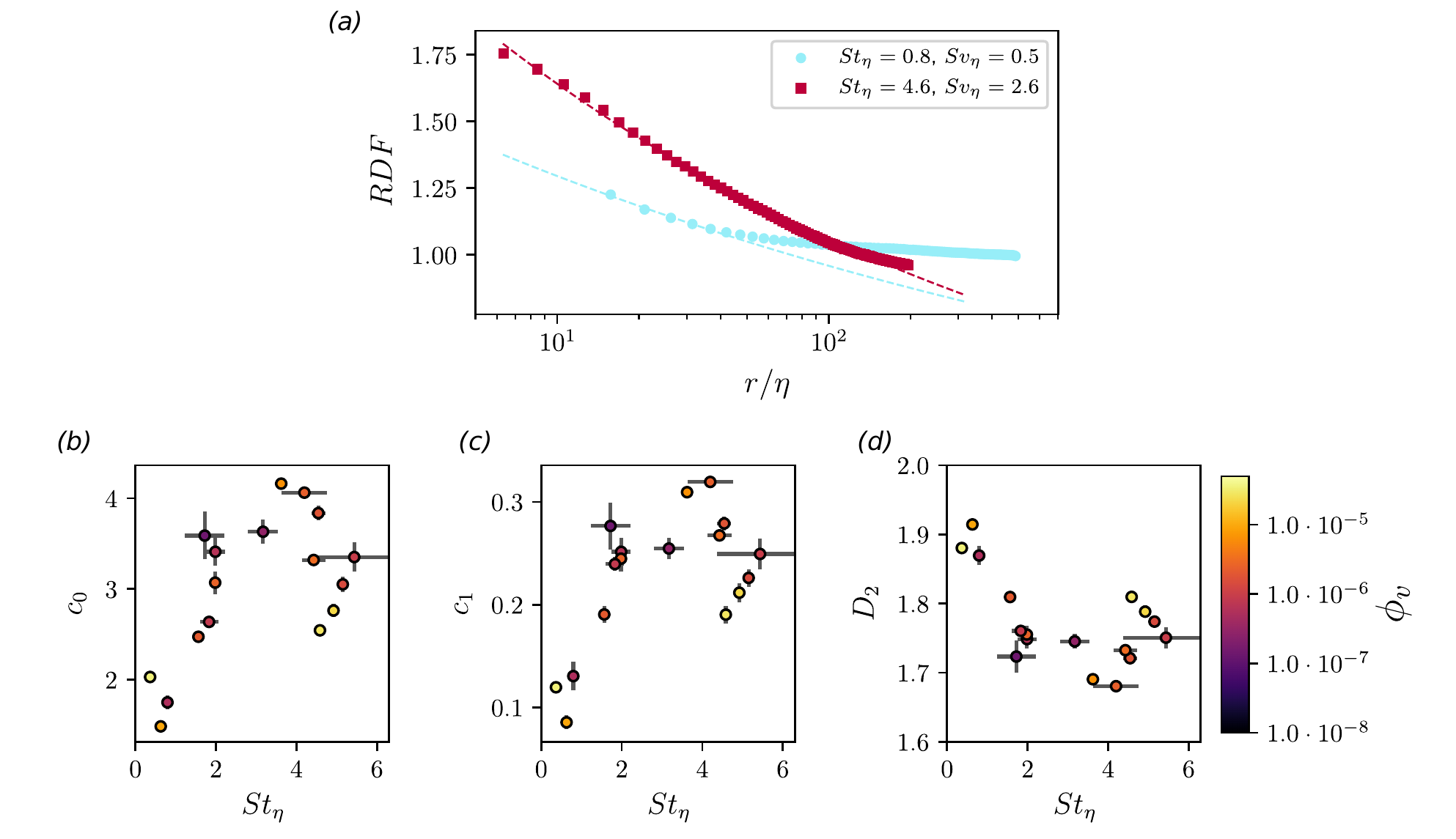}}
		\caption{(a) Example RDFs where symbols represent the calculated RDFs and dashed lines their fits according to equation 3.2. (b) and (c) show the fitted values versus $St_{\eta}$, and (d) shows the value of the correlation dimension.}
		\label{fig:rdf1}
	\end{figure}
\end{center}
\indent The large-FOV measurements allow us to probe the spatial distribution over much greater scales. Figure \ref{fig:rdf2}a clearly indicates that considerable clustering occurs over lengths $O(L)$. For a quantitative assessment, we consider the original power-law model proposed by \citet{ReadeCollins2000}:
\begin{equation}
g(r/\eta) -1 = c_0^*(r/\eta)^{-c_1^*}e^{-c_2^* r / \eta}
\end{equation}
which, unlike (3.2), does recover the return to unity at large separations. The excellent fit to the data over the entire window confirms the observation made by \citet{ReadeCollins2000}, that the RDFs of preferentially concentrated particles have a power-law decay at small scales and an exponential tail at large scales. Figure \ref{fig:rdf2}b,c,d display the least-square-fit coefficients as a function of $St_{\eta}$. The $c_2^*$ coefficient decreases as $St_{\eta}$ increases, implying a greater spatial extent of clustering for the more inertial particles. This may be due to the more inertial particles responding to larger time scales of the turbulence, and to the influence of gravitational settling as mentioned above. The length scale of the large-scale clustering can be estimated from the exponential decay as $\eta/c_2$, which for $St_{\eta} > 1$ is about $300\eta - 400\eta$, close to the integral scale of the turbulence. Taken together, these results confirm that clustering can extend over larger scales for heavier particles. This is in agreement with the conceptual picture of \citet{GotoVassilicos2006} and \citet{YoshimotoGoto2007} and the simulations of \citet{Bec2010} and \citet{IrelandBraggCollins2016a, IrelandBraggCollins2016b}, which showed that particles of $St_{\eta} > 1$ respond to eddies in the inertial range. However, as we will reiterate in the next sub-section, the present results indicate that some level of clustering may extend even beyond, approaching the integral scales.
\begin{center}
	\begin{figure}
		\centerline{\includegraphics[width=\textwidth]{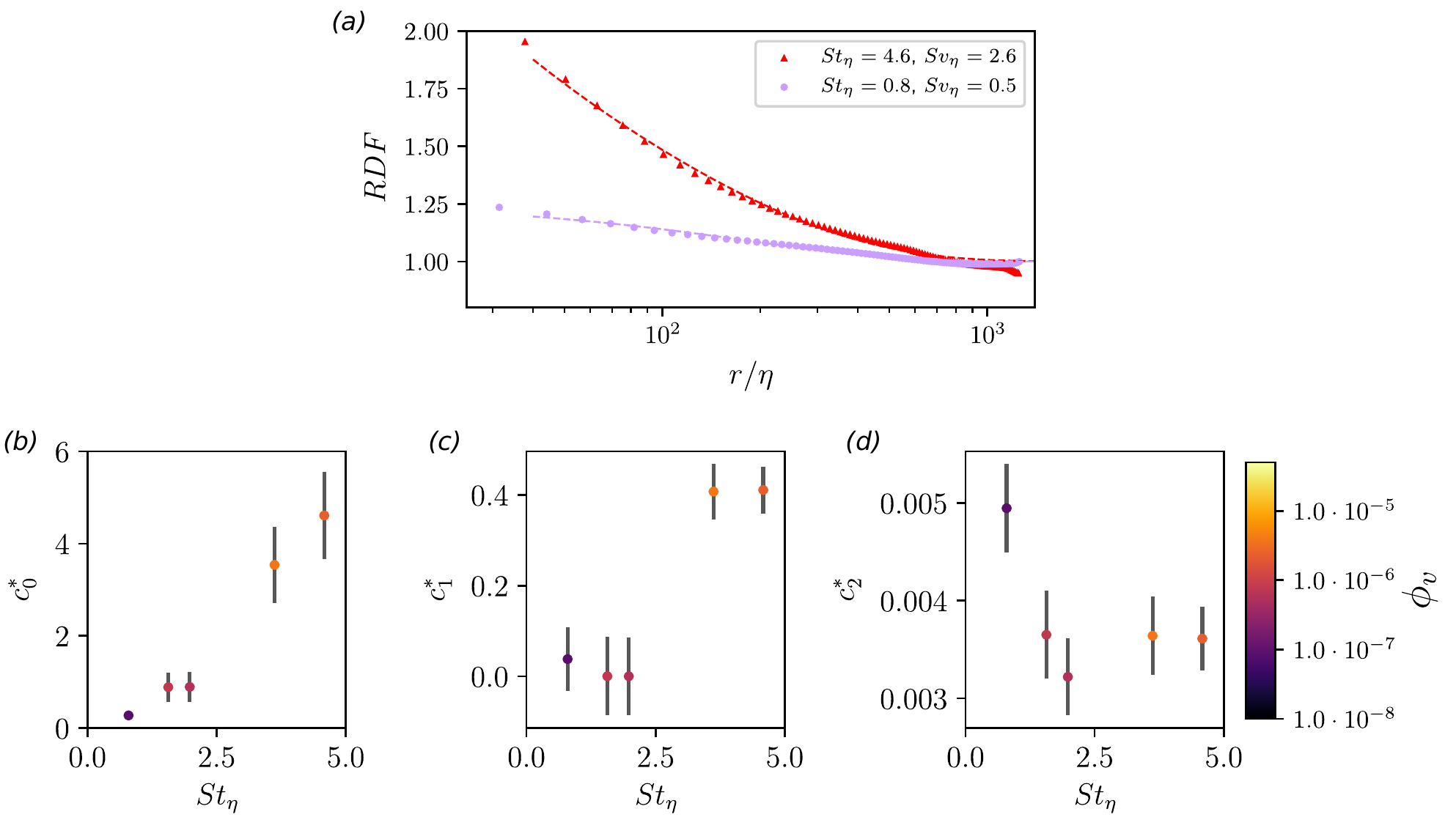}}
		\caption{(a) Example RDFs where symbols represent the calculated RDFs and dashed lines their fits according to equation 3.3. (b), (c) and (d) show the fitted coefficients versus $St_{\eta}$.}
		\label{fig:rdf2}
	\end{figure}
\end{center}\indent The degree of clustering can also be evaluated from the Vorono\"i diagrams. We remark that the type of information provided by this method is somewhat different than the RDF. The latter is strictly a two-particle quantity, while the shape and size of the Voronoi cells result from the mutual position of multiple particles. Therefore, we look for an insight complementary to our RDF results. In figure \ref{fig:clust_quant}a we plot the standard deviation of the Vorono\"i cell areas $\sigma_A$ as a function of the Stokes number, normalizing it by the expected value for particles distributed according to a random Poisson process, $\sigma_{RPP} \approx 0.53$ \citep{Monchaux2010}. As a general trend, clustering is most pronounced for particles of $St_{\eta} \approx 1$, in agreement with previous studies \citep{Monchaux2010, Tagawa2012, DejoanMonchaux2013, DejoanMonchaux2017}. However, the significant scatter suggests that other parameters may also play a role. Indeed, in their grid turbulence study, \citet{Sumbekova2017} found that $\sigma_A$ was strongly dependent on $Re_{\lambda}$, moderately on $\phi_v$, and negligibly on $St_{\eta}$. While the considerable degree of polydispersity in their experiments may have influenced such conclusion, their results convincingly indicated that clustering is affected by a range of turbulent scales, whose breadth is controlled by $Re_{\lambda}$. Moreover, as pointed out by \citet{Baker2017}, $\sigma_A$ is not only a function of the concentration of the clustered particles, but also of the size and distribution of the voids, and the latter are strongly influenced by the inertial and integral scales of the turbulence \citep{YoshimotoGoto2007}.\\
\indent The decrease in $\sigma_A$ for $St_{\eta} > 1$ is mild. Since increasing $St_{\eta}$ also implies increasing $Sv_{\eta}$, this again suggests that, in this range, gravitational settling may enhance clustering. This idea is supported by figure \ref{fig:clust_quant}b, showing the fraction of particles belonging to coherent clusters (according to the definition in \textsection 2.3) plotted versus $\sigma_A$. A clear correlation is visible, indicating that the number of clustered particles, $N_c$, is is similarly affected by the physical parameters. The values are possibly underestimated, because particles in highly concentrated regions are more likely to be overshadowed by neighboring particles and go undetected. Still, the results are in fair agreement with the DNS of \citet{Baker2017}, where less than 3\% of the particles with $St_{\eta} < 1$ belonged to coherent clusters, with the percentage increasing up to 14\% for $St_{\eta} = O(10)$. Therefore, the more inertial (and faster falling) particles are more likely to belong to clusters; or, equivalently, they tend to form clusters that are more numerous, larger, or denser. The question of the cluster size and the concentration within them is addressed in the following section. 
\begin{center}
	\begin{figure}
		\centerline{\includegraphics[width=\textwidth]{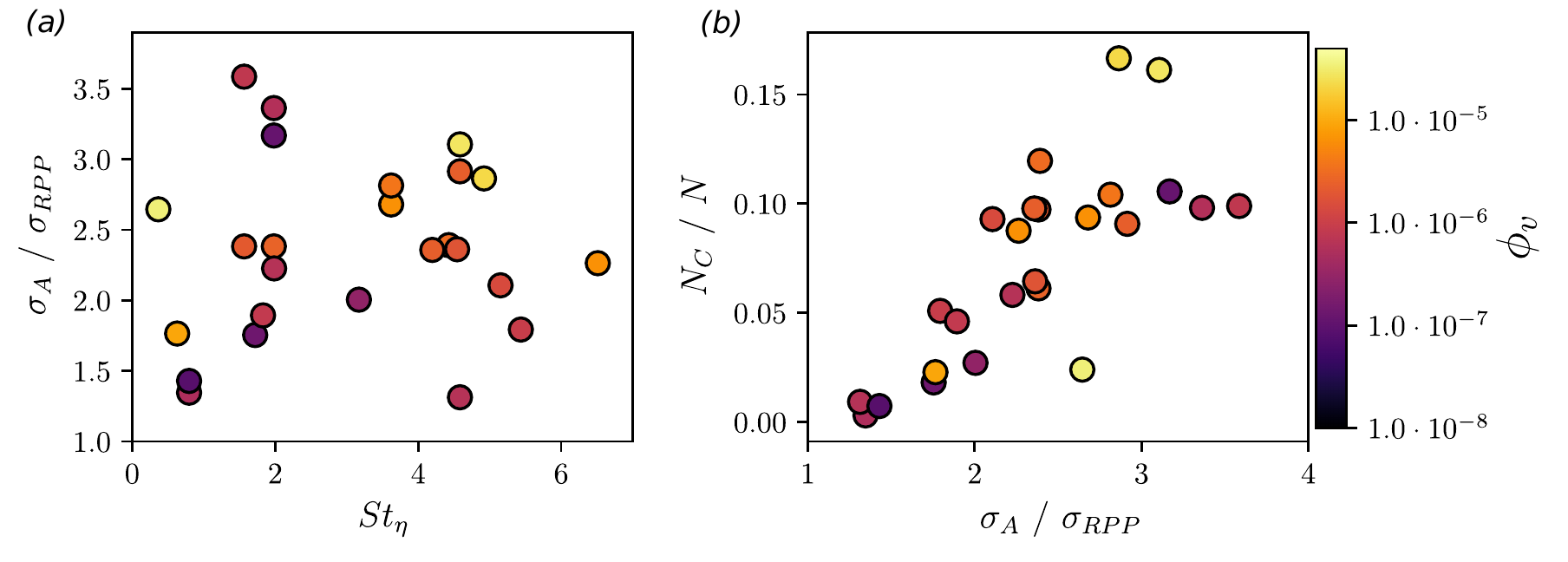}}
		\caption{Two qualifications of clustering intensity using the Vorono\"i tessellation method: (a) $\sigma_{A} / \sigma_{RPP}$ representing departure from a random particle distribution versus Stokes number. (b) Fraction of inertial particles belonging to coherent clusters as a function of normalized $\sigma_A / \sigma_{RPP}$.}
		\label{fig:clust_quant}
	\end{figure}
\end{center}
\subsection{Individual clusters}
\begin{center}
	\begin{figure}
		\centerline{\includegraphics[width=5in]{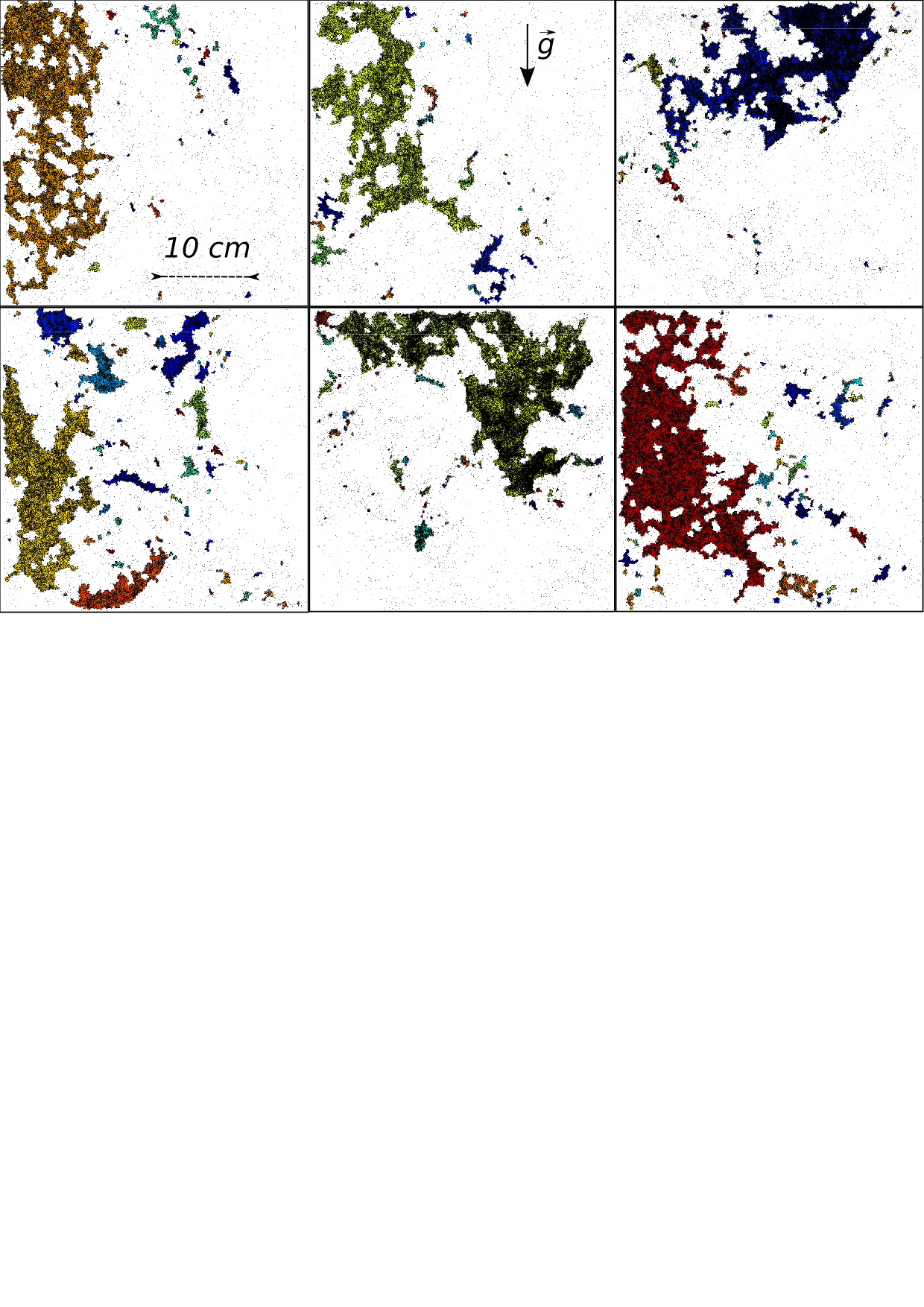}}
		\caption{Example clusters for the cases $St_{\eta} =$ 1.6 and 4.6  imaged in the large FOV, highlighting the wide variety of sizes and shapes. }
		\label{fig:bigboi}
	\end{figure}
\end{center}

\begin{center}
	\begin{figure}
		\centerline{\includegraphics[width=\textwidth]{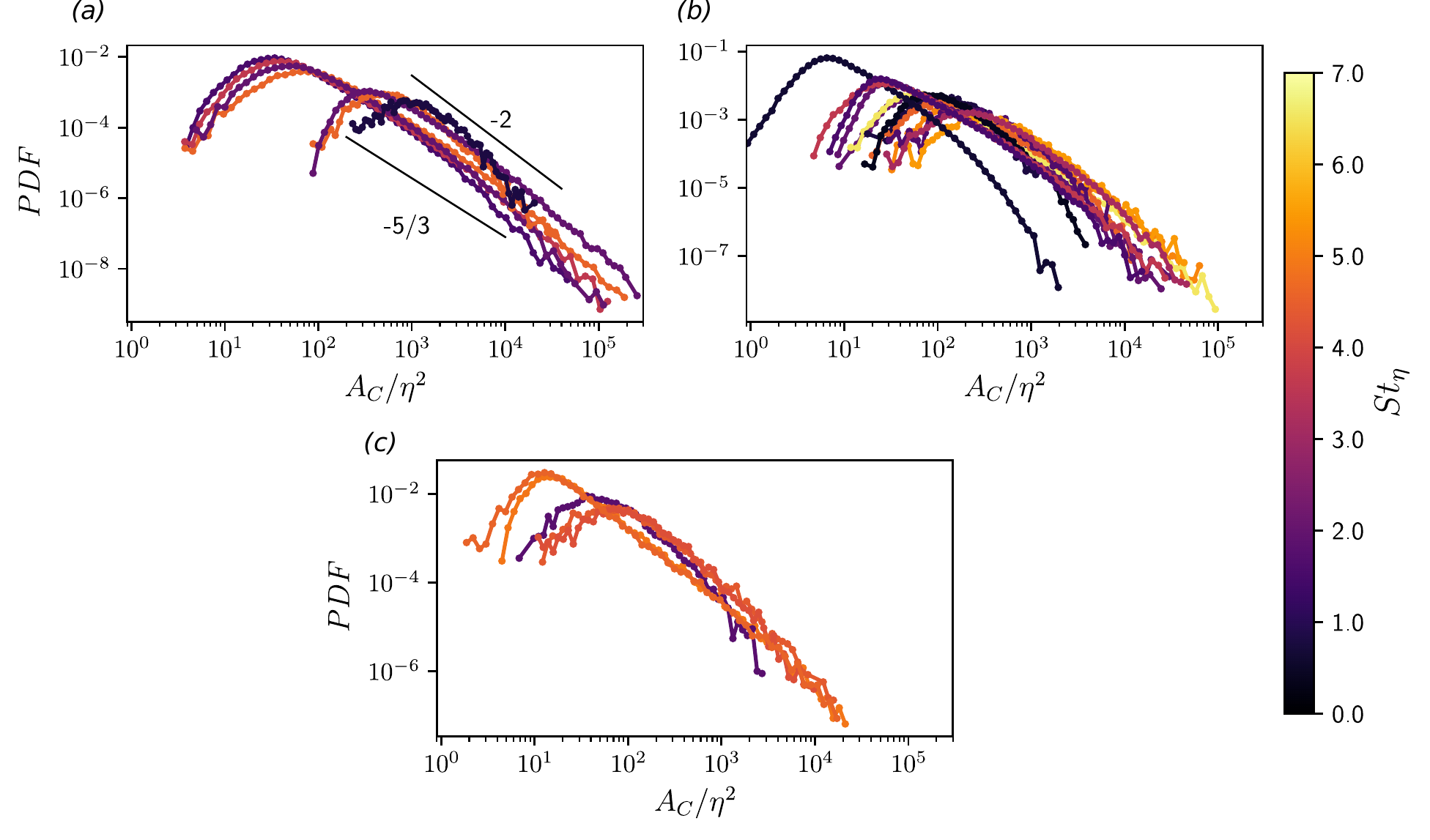}}
		\caption{PDFs of cluster area normalized by $\eta$ and separated by field-of-view size. Largest FOV (~30x30 cm$^2$) in (a), medium field (~14.5x14.5 cm$^2$) in (b), and the smallest field in (c) (~4.5x4.5 cm$^2$).}
		\label{fig:clust_pdfs}
	\end{figure}
\end{center}

\begin{center}
	\begin{figure}
		\centerline{\includegraphics[width=4.5in]{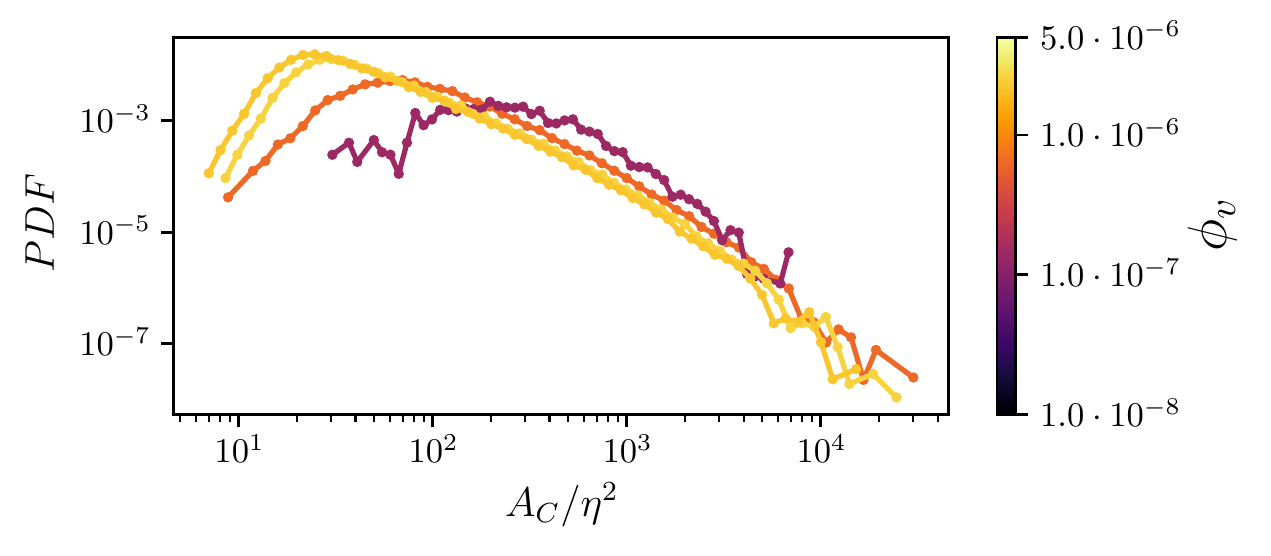}}
		\caption{PDFs of cluster area for experiments with $St_{\eta}$ between 1 and 2.5 (30 $\mu$m glass spheres) in the medium field-of-view showing how increasing volume fraction shifts PDFs to smaller scales.}
		\label{fig:sim_stks}
	\end{figure}
\end{center}

\begin{center}
	\begin{figure}
		\centerline{\includegraphics[width=3.5in]{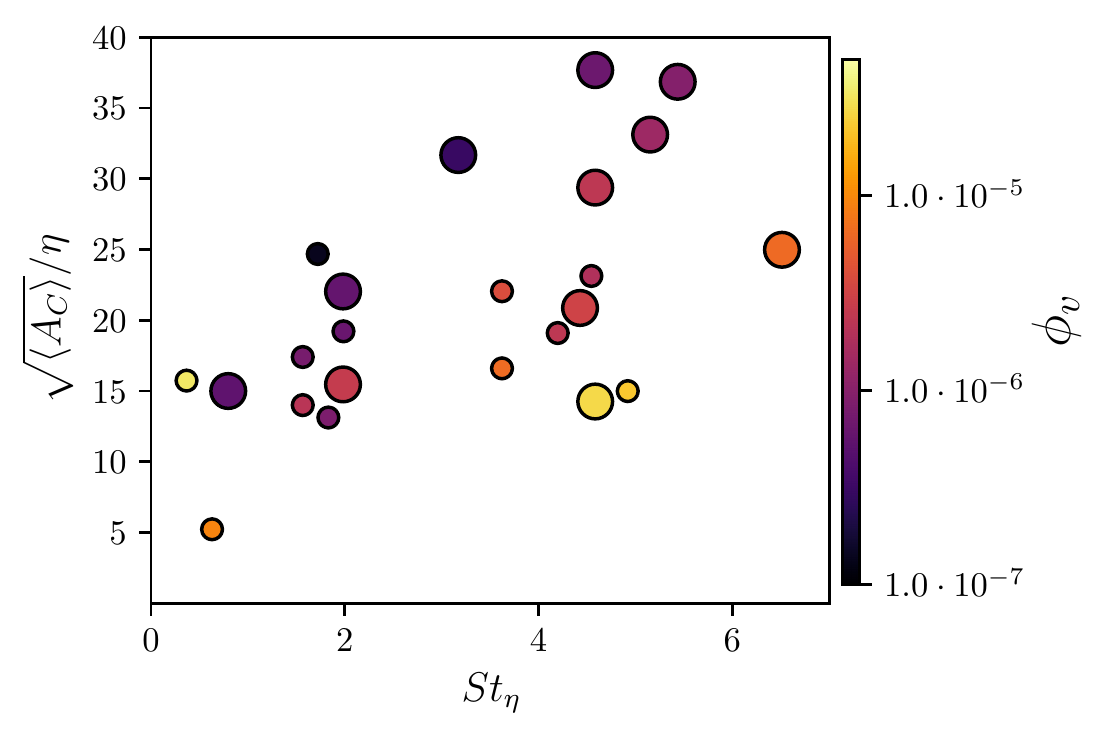}}
		\caption{Mean cluster area versus particle inertia. Marker size is proportional to $Re_{\lambda}$ of the underlying turbulent flow.}
		\label{fig:mean_area}
	\end{figure}
\end{center}

\begin{center}
	\begin{figure}
		\centerline{\includegraphics[width=\textwidth]{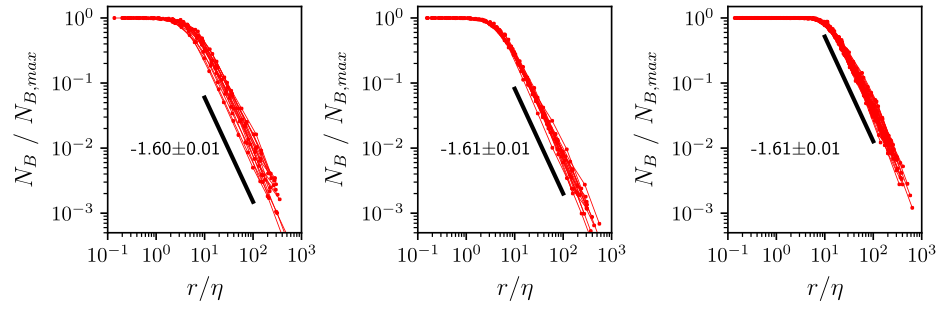}}
		\caption{$N_B(r)$ for three cases with $St_{\eta} = 4.6$ in (a), $3.6$ in (b)) and $2.0$ in (c), normalized by the respective maximum. The box-counting dimension of 1.6 is consistent for all cases.}
		\label{fig:box_count}
	\end{figure}
\end{center}

\begin{center}
	\begin{figure}
		\centerline{\includegraphics[width=\textwidth]{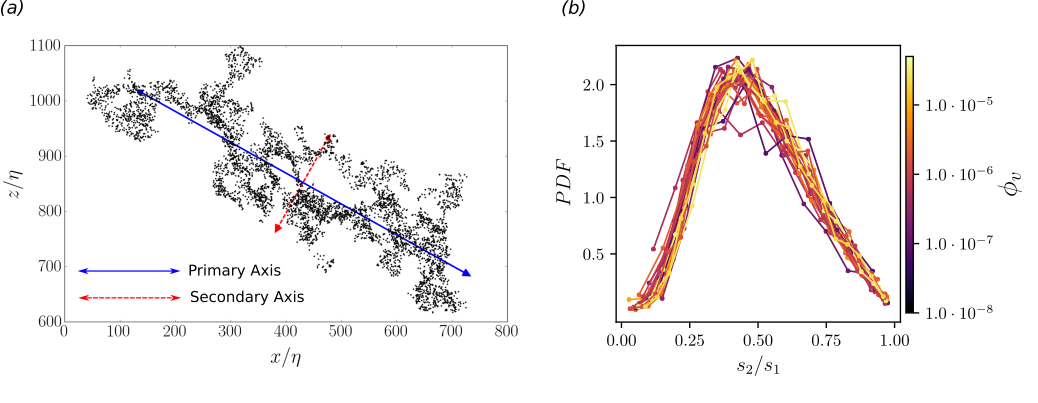}}
		\caption{(a) A single coherent cluster, with primary and secondary axes as computed through SVD. The length of each arrow is proportional to the corresponding singular value. (b) PDFs of the aspect ratios of clusters for each case.}
		\label{fig:aspect_ratio}
	\end{figure}
\end{center}

\begin{center}
	\begin{figure}
		\centerline{\includegraphics[width=\textwidth]{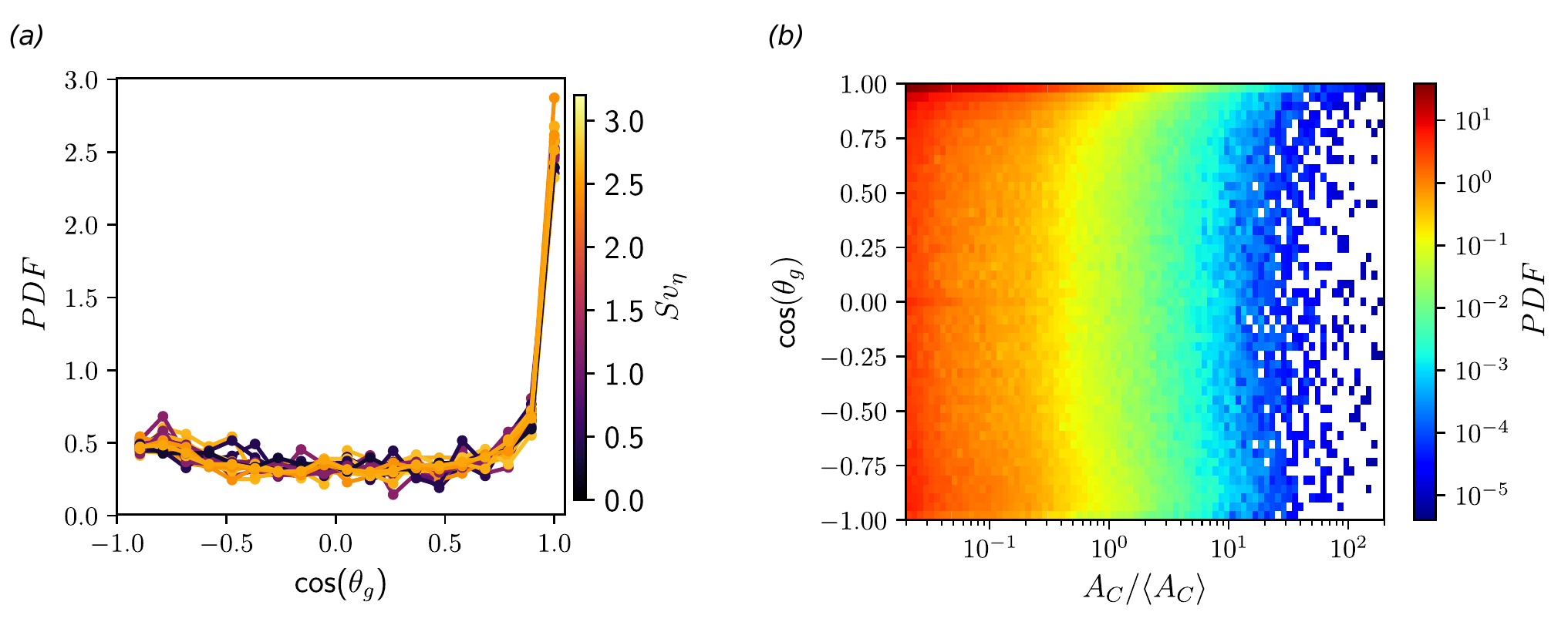}}
		\caption{(a) PDFs of the cosine of the angle between clusters' primary axis and the vertical. (b) Joint PDF of angle from vertical and size of cluster.}
		\label{fig:orientation}
	\end{figure}
\end{center}

\begin{center}
	\begin{figure}
		\centerline{\includegraphics[width=\textwidth]{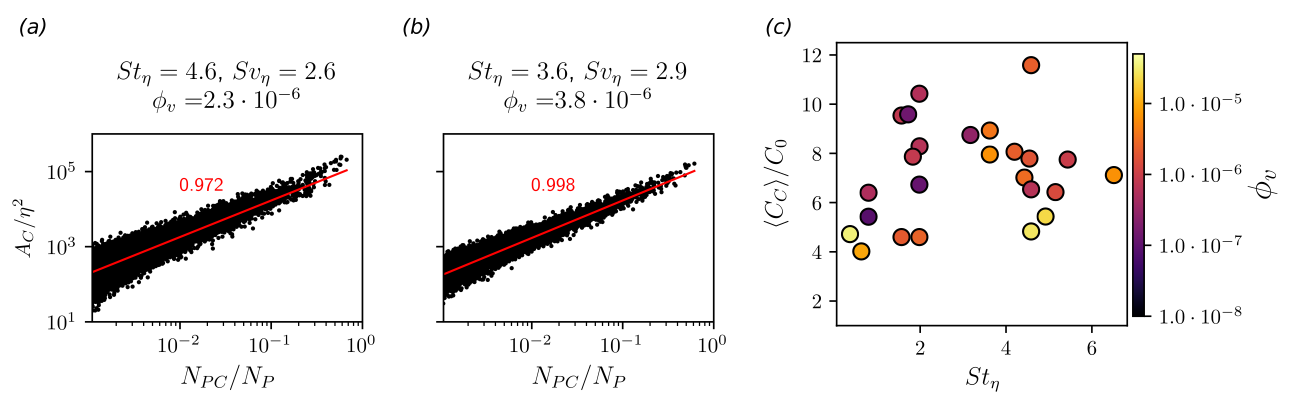}}
		\caption{(a,b): Scatter plots of cluster area (normalized by the Kolmogorov scale) versus the number of particles in each cluster (normalized by the total number of particles in the domain) for two sample cases. (c): Average particle concentration in the clusters, normalized by the global concentration, plotted versus $St_{\eta}$ and colored by overall solid-volume fraction. }
		\label{fig:cluster_conc}
	\end{figure}
\end{center}

\indent The multi-scale nature of the clustering process is reflected in the features of the individual clusters. Figure \ref{fig:bigboi} shows several sample clusters from various instantaneous realizations, as captured by the large-FOV measurements and identified by the Voronoi diagram method, illustrating the variety of sizes and complex shapes of these objects. Some of them are even larger than the integral scales of the flow, often exceeding the limits of the imaging window. Their borders are jagged and convoluted, and their bodies non-simply connected. In the following, we provide quantitative support to these observations. We stress that the objects captured by 2D imaging are cross-section of 3D clusters; this naturally conditions our ability to assess their topology. Such limitation, however, is not expected to overshadow the main conclusions of the analysis. 
\indent Figure \ref{fig:clust_pdfs} shows the PDF of the areas $A_C$ of the clusters, coherent and not, distinguishing between measurements obtained over the small, intermediate, and large FOV. In agreement with \citet{Sumbekova2017} and \citet{Baker2017}, most cases display a power-law behavior over several decades, suggesting a self-similar hierarchy of structures, possibly associated to the scale-invariant properties of the underlying turbulent field \citep{MoisyJimenez2004, GotoVassilicos2006}. The data is consistent with the previously suggested values of $-2$ and $-5/3$ for the power-law exponent for planar measurements \citep{Monchaux2010, Obligado2014, Sumbekova2017}.\\
\indent As expected, the spatial resolution influences the size distributions. The small FOV is affected by a cut-off at large scales. At small scales, the limited resolution in the large FOV makes particles more likely to go undetected due to the glare of their neighbors, reducing the probability of finding small clusters. The latter effect can partly explain why the area threshold for self-similarity (see vertical dashed line in figure \ref{fig:voronoi}d) varies significantly between cases, while this was found to be very consistent in the simulations of Baker \textit{et al.} (2017). Another factor influencing this threshold is the particle volume fraction. Although the value of $A^*$ was shown to be robust to particle sub-sampling \citep{Monchaux2012, Baker2017}, varying the number of particles in the domain results in a shift of the cluster area distribution (figure \ref{fig:sim_stks}), which in turn affects the number of detected coherent clusters above the self-similar threshold. Finally, as $\phi_v$ increases, the possibility of significant two-way coupling effects also increases, which may alter the turbulence structure and consequently the clustering process. This aspect will be discussed in \textsection 5.5. 
\indent A remarkable aspect of the distributions in figure \ref{fig:clust_pdfs} is the non-negligible probability of finding clusters of size comparable to the inertial scales of the turbulence. Since our definition of coherent clusters entails a power-law size distribution, and this is found to have an exponent close to -2, the mean area of the coherent clusters is ill-defined. In order to compare with past studies, we calculate the mean area of all clusters $\langle A_C \rangle$, below and above the self-similarity threshold, and plot its square root in figure \ref{fig:mean_area}. The majority of cases display mean sizes between $10\eta$ and $40\eta$, with a generally increasing trend with $St_{\eta}$. Several previous studies have reported mean cluster sizes around $10\eta$ \citep{Aliseda2002, Wood2005, DejoanMonchaux2013}. Most of these studies, however, considered turbulent flows with relatively low $Re_{\lambda}$ and thus limited scale separation. Recently, \citet{Sumbekova2017} investigated droplets in grid turbulence at $Re_{\lambda}$ approaching $500$, and found cluster size distributions and averages comparable with ours. At large Reynolds numbers the spectrum of temporal scales widens, and particles with a broad range of response times become susceptible to clustering mechanisms \citep{YoshimotoGoto2007}. As noted in \textsection 3.2, the more inertial particles respond to larger eddies and therefore can agglomerate in larger sets. The observed dependence of the cluster size with $St_{\eta}$ is consistent with this view. 
\indent To investigate the degree of self-similarity exhibited by individual clusters, and to provide a descriptor of their complex shape, we calculate their box-counting dimension. This has been widely used to characterize the topology of both turbulent structures \citep{MoisyJimenez2004, LozanoDuran2012, Carter2018} and particle clusters in turbulence \citep{Baker2017}. The domain is partitioned into non-overlapping square boxes of side length $r$, and for each cluster we count the number of boxes $N_B$ containing at least one particle. If $N_B(r)$ follows a power-law, i.e. $N_B \sim r^{-D}$, over a sizable range of scales, the exponent $D$ is taken as the box-counting dimension of the object, which is in turn a measure of its fractal dimension. (Several other definitions of fractal dimension exist, and typically they only coincide for mathematical constructs, \citet{Falconer2004}) Relatively large objects are needed for a robust estimate of $D$ over a wide range of scales, and we thus consider only clusters of area larger than $10^4\eta^2$. Additionally, we neglect clusters touching the image boundary, as their silhouette would include spurious straight segments. Figure \ref{fig:box_count} shows, for three sample cases, $N_B(r)$ normalized by the maximum number of boxes for each cluster (corresponding to the smallest box size, $r = \eta$). For each case, curves for only 20 example clusters are shown for clarity. These reveal a remarkably consistent box-counting dimension $D \approx 1.6$ over at least a decade of scales; the same trend is followed by all other cases. \citet{Baker2017} found $D \approx 1.9$ for 3D clusters. Relating the box-counting dimension of 3D objects and their 2D cross-sections is not straightforward \citep{Tang2006, Carter2018}. Rather, the present result may be compared with that of \citet{Carter2018} who evaluated the box-counting dimension of turbulent coherent structures using 2D PIV in the same facility. They found $D \approx 1.5$, which suggests a strong link between the particle cluster topology and the underlying turbulent flow. Beside the precise value of the box-counting dimension, the main observation is that large clusters of inertial particles do exhibit a scale-invariant shape in the present range of $St_{\eta}$ and $Sv_{\eta}$.
\indent In order to characterize the spatial distribution of particles within each cluster, we use the singular value decomposition (SVD) method introduced by \citet{Baker2017}. The SVD provides the principal axes and corresponding singular values for a particle set. In two dimensions, the primary axis lies along the direction of greatest particle spread from the cluster centroid, the secondary axis being orthogonal to it (figure \ref{fig:aspect_ratio}a). The corresponding singular values $s_1$ and $s_2$ measure the particle spread along the respective principal axes, and can be used as simple shape descriptors through the aspect ratio $s_2$/$s_1$: the limit $s_2$/$s_1$ = 0 corresponds to particles arranged in a straight line, whereas $s_2$/$s_1$ = 1 corresponds to a perfect circle. The PDF of the aspect ratio for all considered cases (figure \ref{fig:aspect_ratio}b) shows that clusters are likely to exhibit aspect ratios between 0.4 and 0.5, reflecting a tendency to form somewhat elongated objects. Furthermore, the distribution has a positive skew, indicating that globular shapes are more common than extremely long streaks. While these observations are influenced by the 2D nature of the technique, they are consistent with the results of \citet{Baker2017} who found that 3D clusters had $s_2$/$s_1$ distributions peaking around 0.5, and were positively skewed. 
\indent The orientation of the primary axis from the SVD analysis provides information on the cluster orientation in space. In figure \ref{fig:orientation}a we plot the PDF of the cosine of $\theta_g$, i.e. the angle between the cluster primary axis and the vertical, evidencing a strong preference for the clusters to align with gravity. That particles tend to agglomerate along their falling direction was found in past one-way coupled simulations \citep{Woittiez2009, DejoanMonchaux2013, Bec2014, IrelandBraggCollins2016b, Baker2017}, indicating the mechanism is not necessarily related to the particle backreaction on the flow. \citet{Baker2017} reasoned that, especially for cases with high $St_{\eta}$ and high $Sv_{\eta}$, particles are influenced by intermittent downward gusts that add to their fallspeed, channeling them and creating elongated quasi-vertical structures. The joint probability distribution of cos($\theta_g$) and $A_C$ (figure \ref{fig:orientation}b) supports this view, showing that a vertical alignment corresponds to generally larger clusters. Further studies, possibly including time-resolved information, are needed to gain a mechanistic understanding of the cluster formation process.
\indent We finally consider the concentration of particles within each coherent cluster, $C_C = N_{PC}/A_C$, where $N_{PC}$ is the number of particles in each cluster. Figure \ref{fig:cluster_conc}a,b shows scatter plots of cluster areas and number of particles for two representative cases. The excellent fit using a power law of exponent close to unity indicates that the relationship is approximately linear, i.e. the concentration within each cluster is approximately the same for a given case. This trend is recovered for all cases. Considering the wide range of sizes, this result (reported by \citet{Baker2017} at a much lower $Re_{\lambda}$) indicates again that the clusters display scale-invariant features. Figure 14c illustrates the average in-cluster concentration as a function of $St_{\eta}$. Despite the scatter (which points to the concurrent effect of the multiple parameters at play), one notices an increase up to $St_{\eta} \approx 2$, followed by a plateau. The concentration within clusters can be up to an order of magnitude higher than the average over the whole particle field ($C_0 = 1/\langle A\rangle$); these values are likely underestimated as particles may shadow each other at high local concentration. The present results are comparable to those from the experiments by \citet{Monchaux2010} and the DNS by \citet{Baker2017}.
\section{Settling velocity}
\subsection{Mean settling velocity}
\begin{center}
	\begin{figure}
		\centerline{\includegraphics[width=\textwidth]{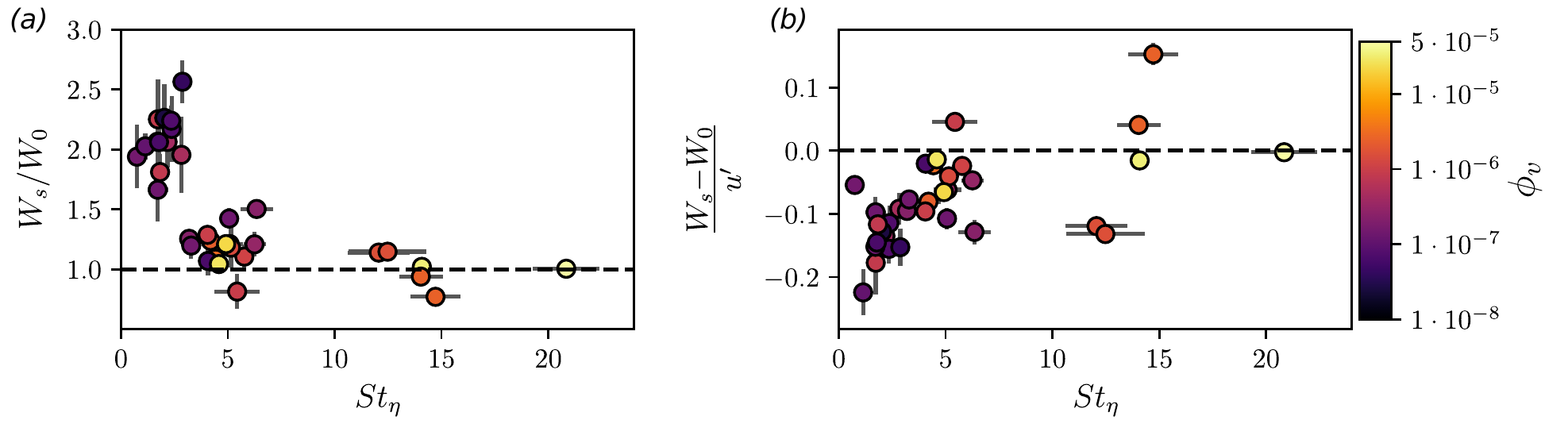}}
		\caption{(a) Measured settling velocity normalized by the still-fluid Stokes value versus $St_{\eta}$ and colored by solid-volume fraction. (b) Measured settling velocity variation from the still-fluid Stoked value normalized by the fluid r.m.s. fluctuations.}
		\label{fig:settling1}
	\end{figure}
\end{center}

\begin{center}
	\begin{figure}
		\centerline{\includegraphics[width=\textwidth]{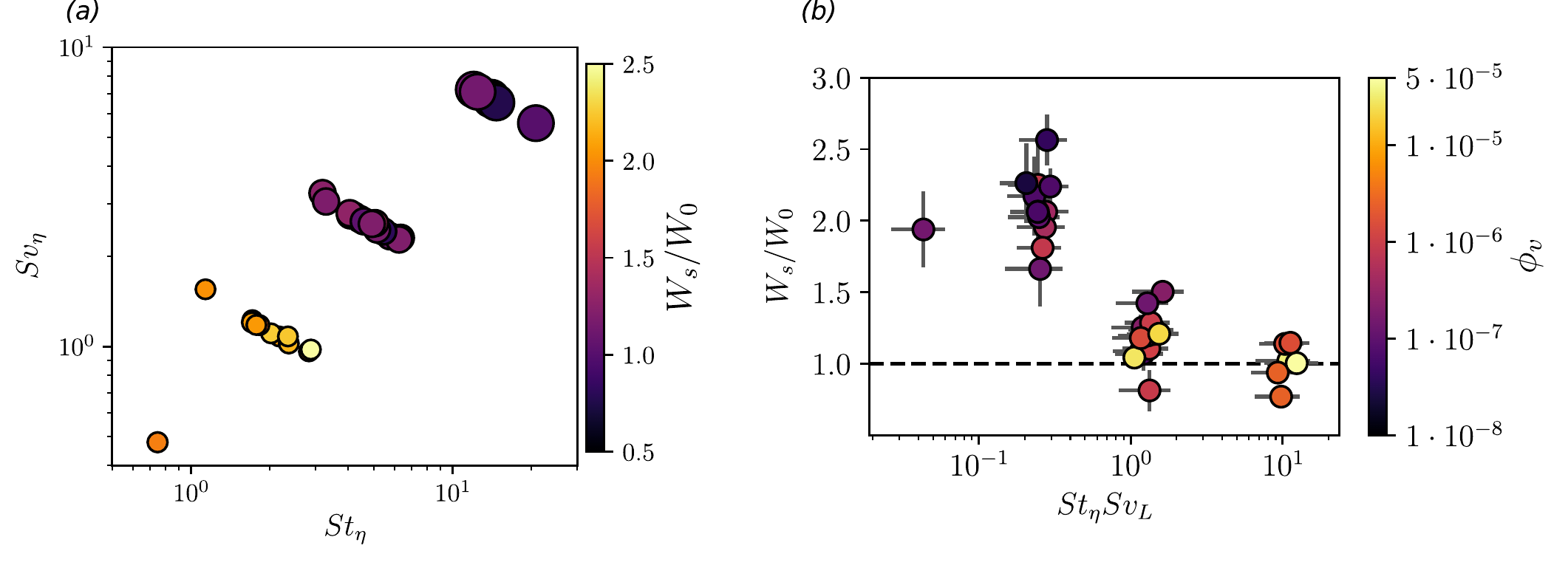}}
		\caption{(a) Settling increase plotted as a function of both $St_{\eta}$ and $Sv_{\eta}$. (b) Settling increase as a function of the mixed scale, $St_{\eta}Sv_L$.}
		\label{fig:mixed_settle}
	\end{figure}
\end{center}

In this section we present and discuss the settling velocity measurements $W_s$, obtained ensemble-averaging over all particles and realizations for each case. In figure \ref{fig:settling1}a this is normalized by $W_0$ (so that values greater and smaller than one indicate turbulence-enhanced and turbulence-inhibited settling, respectively) and plotted against $St_{\eta}$. The main contribution to the error is the uncertainty in $\tau_p$ due to the particle size variance, and the non-zero vertical air velocity measured at the same time as the settling. We only plot cases in which the mean vertical fluid velocity is smaller than $0.25W_s$, and in fact in most cases it is $0.01 - 0.05W_s$. The first observation from the plot is that the vast majority of cases display strong settling enhancement, especially for $St_{\eta} \approx 1$, which is consistent with most previous numerical \citep{WangMaxey1993, Bosse2006, DejoanMonchaux2013, Bec2014, IrelandBraggCollins2016b, Rosa2016} and experimental studies \citep{Aliseda2002, YangShy2003, YangShy2005, Good2014}. The amount of such increase is more remarkable, with the settling velocity being enhanced by a factor 2.6 for $St_{\eta} \approx 1 - 2$. As mentioned in the Introduction, most numerical studies reported maximum increase in fallspeed between about 10\% and 90\%; some experiments \citep{YangShy2003, YangShy2005} found even smaller values. The present results instead indicate that turbulence can lead to a multi-fold increase in settling rate, which agrees with the conclusions from the field study of \citet{Nemes2017}.\\
\indent Similar levels of settling enhancement could also be deduced from the data of \citet{Aliseda2002} and \citet{Good2014}; and while the former used concentrations where collective effects are expected ($\phi_v \geq 10^{-5}$), the latter used particle loadings small enough to neglect two-way coupling ($\phi_v \approx 10^{-6}$). In fact, those authors did not explicitly mention a multi-fold increase in vertical velocity, as they mostly plotted their data as $(W_s-W_0)/u^{\prime}$. We present this scaling in figure \ref{fig:settling1}b, where a maximum settling enhancement of $0.28u^{\prime}$ is found for $St_{\eta} \approx 1$, again in good agreement with those authors. (Note that the vertical velocity is positive when downward, hence negative values imply settling enhancement and vice versa.) The scatter and the superposition of multiple factors prevent distinguishing a clear trend at the larger $St_{\eta}$. Those data points are also at relatively high $\phi_v$, which may have a non-trivial influence on settling, as we discuss later. Thus, the reduced settling exhibited by some of the most inertial cases, while it might appear consistent with recent results \citep{Good2014, Rosa2016}, should be considered with caution. 
\indent\indent Figure \ref{fig:mixed_settle} shows the results in the $St_{\eta}-Sv_{\eta}$ plane. This provides a clearer view of the data, as both parameters are expected to have significant influence on the dynamics. The maximum enhancement of settling rate occurs when both $St_{\eta}$ and $Sv_{\eta}$ are close to unity, in broad agreement with \citet{Good2014} and \citet{Rosa2016}. The distribution of values suggests that a dependence with $StSv$ may capture the observed trend. Figure \ref{fig:mixed_settle}b shows the settling enhancement ratio against the group $St_{\eta}Sv_L$, displaying a significantly improved collapse of the data. This scaling follows the argument of \citet{Nemes2017} that $\tau_{\eta}$ and $u^{\prime}$ are the main time and velocity scales, respectively, determining increase of fallspeed by turbulence. That both the small and large eddies impact the settling process has been acknowledged \citep{Good2014}, and already \citet{WangMaxey1993} favored $u^{\prime}$ over $u_{\eta}$ as driving parameter. \cite{YangLei1998} explicitly indicated $\tau_{\eta}$ and $u^{\prime}$ as the correct flow scales, reasoning that the former controlled clustering and the latter controlled the drag experienced by the particles. The group $St_{\eta}Sv_L$ can be interpreted as the ratio of the particle stopping distance ($\tau_p^2g$) and a mixed length scale ($\tau_{\eta}u^{\prime}$); settling enhancement appears most effective when this ratio is $O(0.1)$. This is approximately the condition at which \citet{Nemes2017} reported turbulence-augmented fallspeeds of snowflakes in the atmospheric surface layer ($Re_{\lambda} \approx 10^3$). Mixed-scaling arguments have been successfully used in various turbulent flows (e.g., in boundary layers, \citep{DeGraff2000}) but their theoretical underpinning poses issues which are beyond the scope of the present study.
\indent\indent Overall, the results presented in this section indicate that turbulence greatly enhances the settling velocity of sub-Kolmogorov particles with Stokes number around unity, which is consistent with the preferential sweeping mechanism proposed by \citet{WangMaxey1993}. However, a full demonstration of this view requires the simultaneous measurements of particle and fluid velocity. These will be presented in \textsection 5.2.
\subsection{Settling velocity conditioned on particle concentration}
\begin{center}
	\begin{figure}
		\centerline{\includegraphics[width=0.75\textwidth]{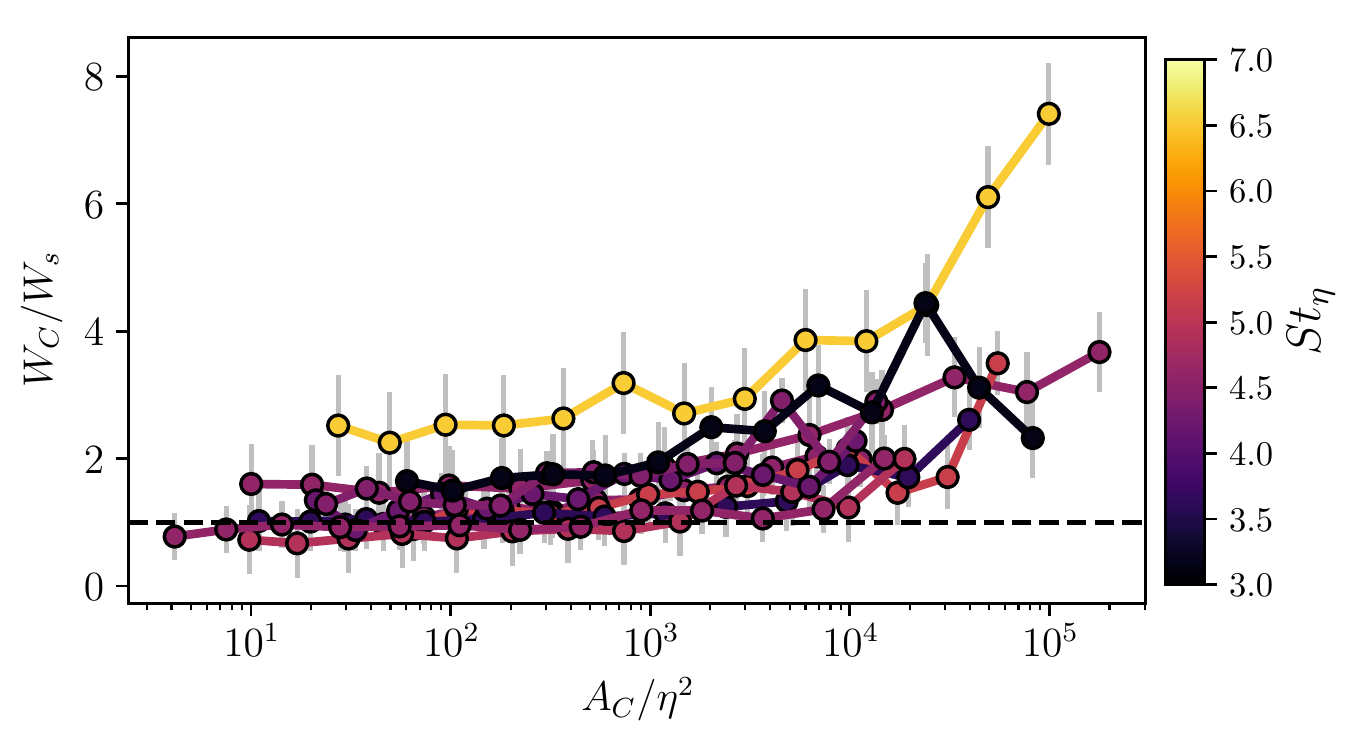}}
		\caption{Average settling velocity of all clusters as a function of cluster size. Error bars indicate $\pm$ the standard deviation.}
		\label{fig:clust_settle}
	\end{figure}
\end{center}

\begin{center}
	\begin{figure}
		\centerline{\includegraphics[width=0.75\textwidth]{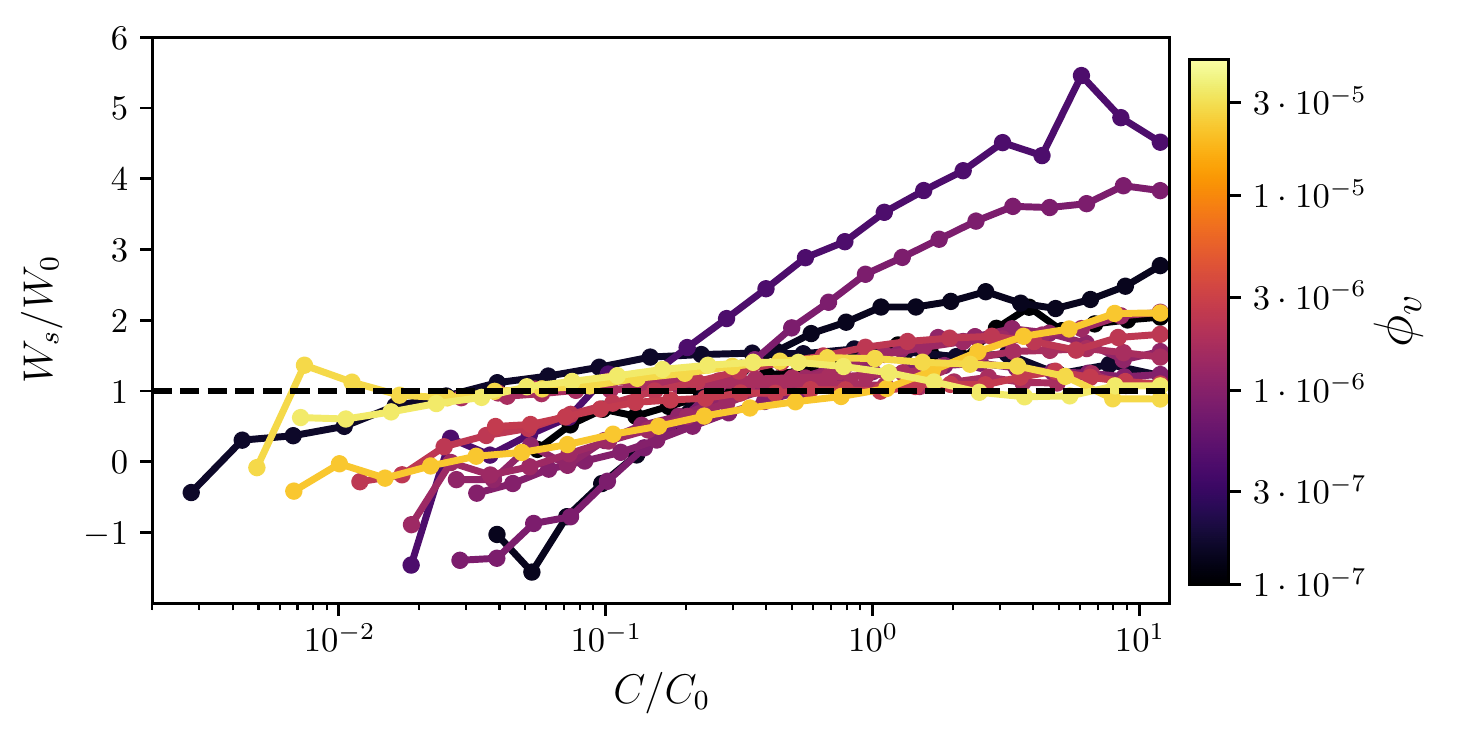}}
		\caption{Normalized vertical particle velocity conditioned on the local particle concentration.}
		\label{fig:conc_settle}
	\end{figure}
\end{center}

To explore the interplay between the particle accumulation and settling mechanisms, we consider the fallspeed associated to individual coherent clusters. In figure \ref{fig:clust_settle} we plot the cluster settling velocity $W_C$, obtained by averaging the vertical velocity of all particles belonging to a given clustered set. This is normalized by the mean settling velocity $W_s$, and plotted against the cluster area. Overall, clusters settle significantly faster than the mean, and there is an apparent trend of increasing fallspeed with cluster size, especially for the larger objects. There are two possible interpretations for this result. On one hand, clustered particles may affect the flow by virtue of their elevated concentration, exerting a “collective drag” on the surrounding fluid that results in increased settling velocity. This view reflects the argument proposed by \citet{Bosse2006} in interpreting their two-way coupled DNS study. On the other hand, particles may be merely oversampling downward regions of flow according to the preferential sweeping mechanism, and therefore cluster in such regions, leading to the observed trend. This latter interpretation, which does not require any significant two-way coupling between the dispersed and continuous phase, is consistent with the results of \citet{Baker2017}, who reported cluster fallspeeds up to twice the mean particle settling velocity in their one-way-coupled DNS.
\indent\indent Contrasting the effect of local and global concentration may provide further hints. In figure \ref{fig:conc_settle}, the particle settling velocity of all particles $W_s$ (normalized by the still-air fallspeed $W_0$) is plotted against the local relative concentration $C/C_0$ (which is readily available for each particle from the Vorono\"i diagrams). As expected, $W_s/W_0$ increases monotonically with $C/C_0$, in agreement with the trends reported by \citet{Aliseda2002}. Indeed, particles residing in regions of low concentration are often associated with upward velocity. We also observe, although with some scatter, the beginning of a plateau in the settling enhancement around $C/C_0 \approx 5$. More importantly, the plot clearly indicates that the fallspeed dependence with concentration is strongly mitigated at larger global volume fractions, $\phi_v$. If the high fallspeed of the clusters was mainly due to a collective effect of the particles on the fluid, we would expect such speed to be further enhanced with increasing $\phi_v$. The fact that the opposite is true rather suggests that the augmented cluster settling is mainly caused by preferential sweeping (or other mechanisms not depending on the mass loading). In fact, figure 18 suggests that two-way coupling may be significant over the considered range of $\phi_v$, but its effect may be subtle: if the particles are altering the turbulence structure, this backreaction can have a non-trivial effect on the settling rate. In general, it should be remarked that the simultaneous variation of multiple physical parameters between the considered cases (in this as in other studies) is a confounding factor in determining the role of two-way coupling, and one cannot rule out the influence of collective drag on the settling velocity (as argued by \citet{Huck2018}). Future dedicated studies, in which the global volume fraction is systematically varied while keeping all other parameters constant, may help shed light on this point.
\section{Analysis of simultaneous particle and fluid fields}
In this section we investigate the particle-fluid interaction by exploiting the concurrent PIV/PTV measurements of both phases. These allow us to demonstrate and quantify effects which, although considered hallmarks of particle-laden turbulence, had rarely (if ever) been documented in experiments.
\subsection{Preferential concentration}
\begin{center}
	\begin{figure}
		\centerline{\includegraphics[width=3in]{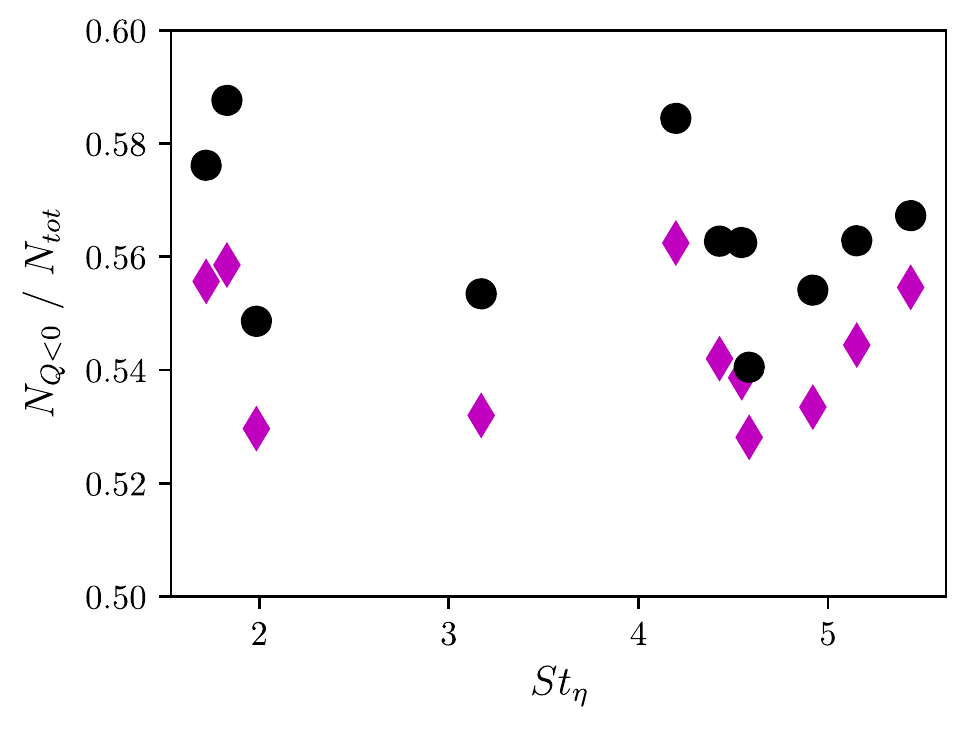}}
		\caption{Fraction of particles in rotation-dominated regions. Black circles represent all particles in the field, while the purple diamonds represent only particles in coherent clusters.}
		\label{fig:Q_per}
	\end{figure}
\end{center}

The fact that inertial particles oversample high-strain/low-vorticity regions, as theorized by \citet{Maxey87} and demonstrated numerically by \citet{SquiresEaton1991a}, was confirmed by several later DNS studies of homogeneous turbulence, at least for $St_{\eta} \leq 1$ \citep{Chun2005, Bec2006, Cencini2006, Coleman2009, SalazarCollins2012, IrelandBraggCollins2016a, Esmaily2016, Baker2017}. To our knowledge, this prediction has not been directly verified by experiments in fully turbulent flows. Indeed, most previous laboratory studies on this topic only captured the dispersed phase \citep{Fessler1994, Aliseda2002, Wood2005, Salazar2008, Saw2008, Gibert2012}, and as such could only provide results consistent with a certain picture of preferential concentration, rather than demonstrating it. 
We characterize the local balance of strain-rate versus rotation in the particle-laden air flow measured by PIV, using the second invariant of the velocity gradient tensor $Q = 1/2(\bm{\Omega}^2-\mathbf{S}^2)$, where $\mathbf{S}$ and $\bm{\Omega}$ are the symmetric and anti-symmetric parts of the velocity gradient tensor \citep{Hunt1988}. To this end, we calculate spatial derivatives using a second-order central difference scheme on the small and medium-FOV fields, where our resolution is sufficient to capture the Kolmogorov scales \citep{Worth2010, Hearst2012}. From the planar data we can only determine the four components in the upper-left 2 x 2 block of the full 3 x 3 velocity gradient tensor. This limitation needs to be kept in mind, because 2D sections of 3D flows can sometimes be misleading \citep{PerryChong94}. However, several studies showed how high-resolution 2D imaging of homogeneous turbulence yields features of the coherent structures and high-order statistics in quantitative agreement with 3D imaging and DNS \citep{Fiscaletti2014, Carter2018, Saw2018}. Therefore, we do not expect the qualitative results of the present analysis to be biased by the nature of the measurements.\\
\indent Figure \ref{fig:Q_per} shows the fraction of inertial particles found in regions where $Q < 0$. As expected, this fraction is larger than 50\% for all cases, confirming that the particles are more likely to be found in strain-dominated regions than rotation-dominated ones. The figure also presents the percentage of clustered particles found in $Q < 0$ regions. Interestingly, the fraction is systematically lower compared to the entire particle set. This suggests that the preferential sampling of high-strain regions might not be the main factor (or at least not the only one) for the formation of clusters over the considered parameter space. Indeed, most of our cases feature particles with $St_{\eta} > 1$, and in this regime several numerical studies indicate that the nature of the clustering mechanism is different compared to weakly inertial particles \citep{BecPRL2007, Coleman2009, BraggCollins2014a, Bragg2015}. In particular, \citet{BraggCollins2014a} argued for the importance of path-history effects (i.e., particles retaining memory of the velocity fluctuations they experienced), while Vassilicos \& coworkers \citep{ChenGotoVas2006, GotoVassilicos2008} proposed that clustering in this range is due to a sweep-stick mechanism (i.e., particles sticking to zero-acceleration points which are swept and clustered by large-scale motions). A critical discussion of these and other possible explanations is beyond the scope of this work. In fact, while instantaneous realizations and velocity statistics may provide support to a given theory \citep[see][]{Obligado2014, Sumbekova2017}, time-resolved measurements would be better suited to inform a mechanistic understanding of the process.
\subsection{Preferential sweeping}
As discussed in \textsection 1 and 4.1, preferential sweeping is considered the most impactful mechanism by which turbulence affects the fallspeed of sub-Kolmogorov particles. Its main manifestation is the tendency of particles with Stokes number of order one to oversample regions of downward velocity fluctuations. This was first theorized by \citet{Maxey1986} and \citet{Maxey87}, demonstrated numerically by \citet{WangMaxey1993}, and confirmed by several other analytical and computational studies \citep{YangLei1998, Davila2001, DejoanMonchaux2013, Frankel2016, Baker2017}. While laboratory studies \citep{Aliseda2002, YangShy2003,YangShy2005, Good2014}and field observations \citep{Nemes2017} showed results consistent with this picture, no direct experimental verification has been reported. Similar to preferential concentration, the challenges associated to two-phase measurements may be responsible.
\begin{center}
	\begin{figure}
		\centerline{\includegraphics[width=3.5in]{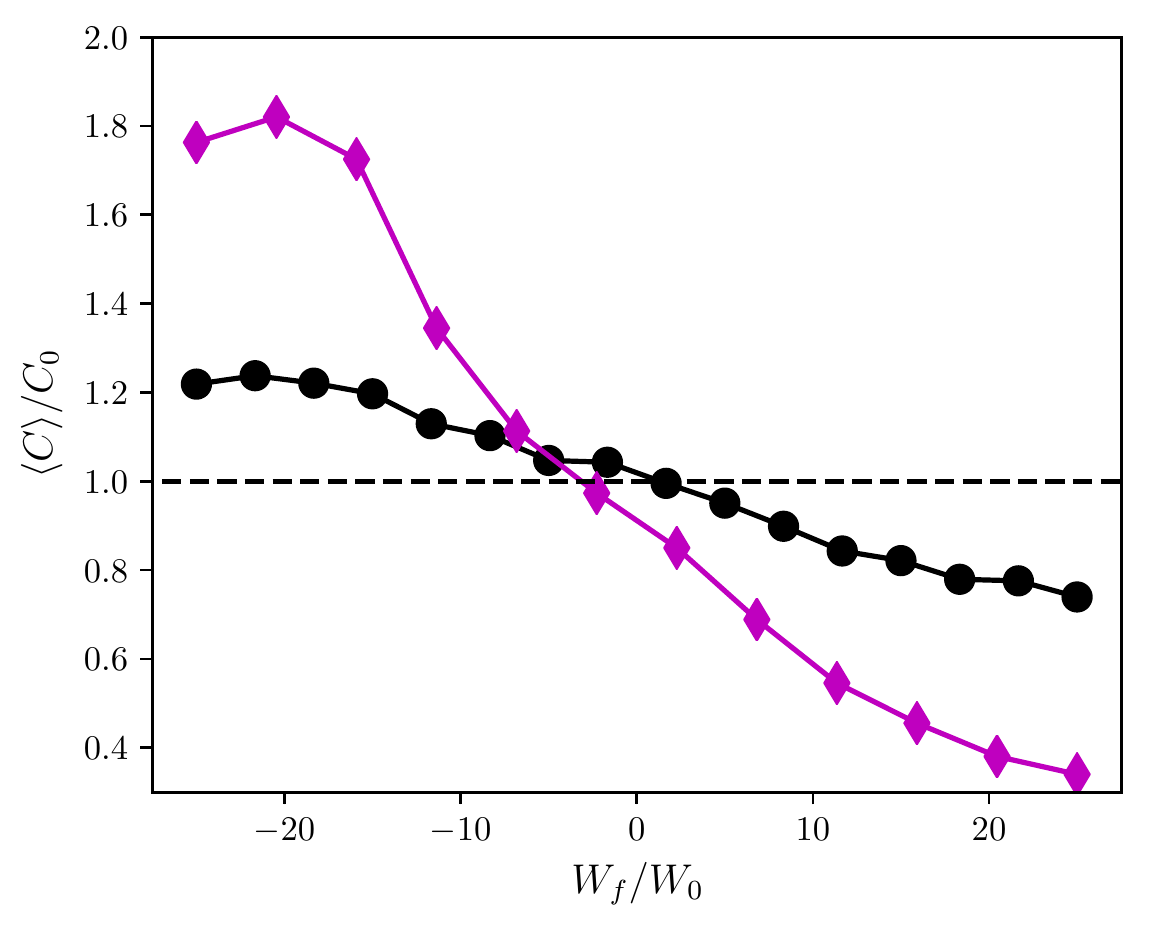}}
		\caption{Example case of the relative particle concentration conditioned on the vertical fluid velocity (normalized by the still-air settling velocity.) Black circles indicate the full particle set, while purple diamonds represent only particles in clusters.}
		\label{fig:sweep}
	\end{figure}
\end{center}
\indent\indent We provide such verification first by considering the particle concentration conditionally averaged on the local fluid velocity. This is obtained by counting the number of inertial particles in each PIV interrogation window, and binning the results by the value of $W_f/W_0$ (because the mean vertical fluid velocity is negligibly small, total and fluctuating components coincide). The relative concentration is calculated as the number of particles in each bin, divided by the sum of window areas associated to that bin, and finally normalized by the global concentration. The procedure is equivalent to that originally adopted by \citet{WangMaxey1993} and later by \citet{Baker2017} to analyze DNS results, with the PIV interrogation windows in lieu of the computational cells. In figure \ref{fig:sweep} we show the result for a representative case, clearly indicating downward fluid velocity corresponding to higher local concentration. When the process is repeated only considering particles belonging to clusters, the trend is significantly more pronounced. This is consistent with the result that clusters fall at faster speeds than the rest of the particles (figure \ref{fig:clust_settle}). At the same time, it also supports the idea that preferential sweeping plays an important role in the clustering of settling particles.\\
\indent To quantify the impact on the settling rate, we consider the vertical component of the fluid velocity at the particle location, $W_f(\mathbf{x_p})$ (figure \ref{fig:interp}). The latter is approximated via a piecewise linear interpolant between the particle position and the four closest fluid velocity vectors; tests with other schemes indicate only a weak dependence with the interpolation method. Error analysis based on the fluid velocity gradient statistics \citep[see][]{Carter2016, CarterColetti2017} yields uncertainty on $W_f(\mathbf{x_p})$ around $2-7$\% $u^{\prime}$. Despite some scatter partly attributable to the several factors at play, the results indicate that preferential sweeping is important for most considered regimes, being the strongest for $St_{\eta} = O(1)$.
\begin{center}
	\begin{figure}
		\centerline{\includegraphics[width=3.5in]{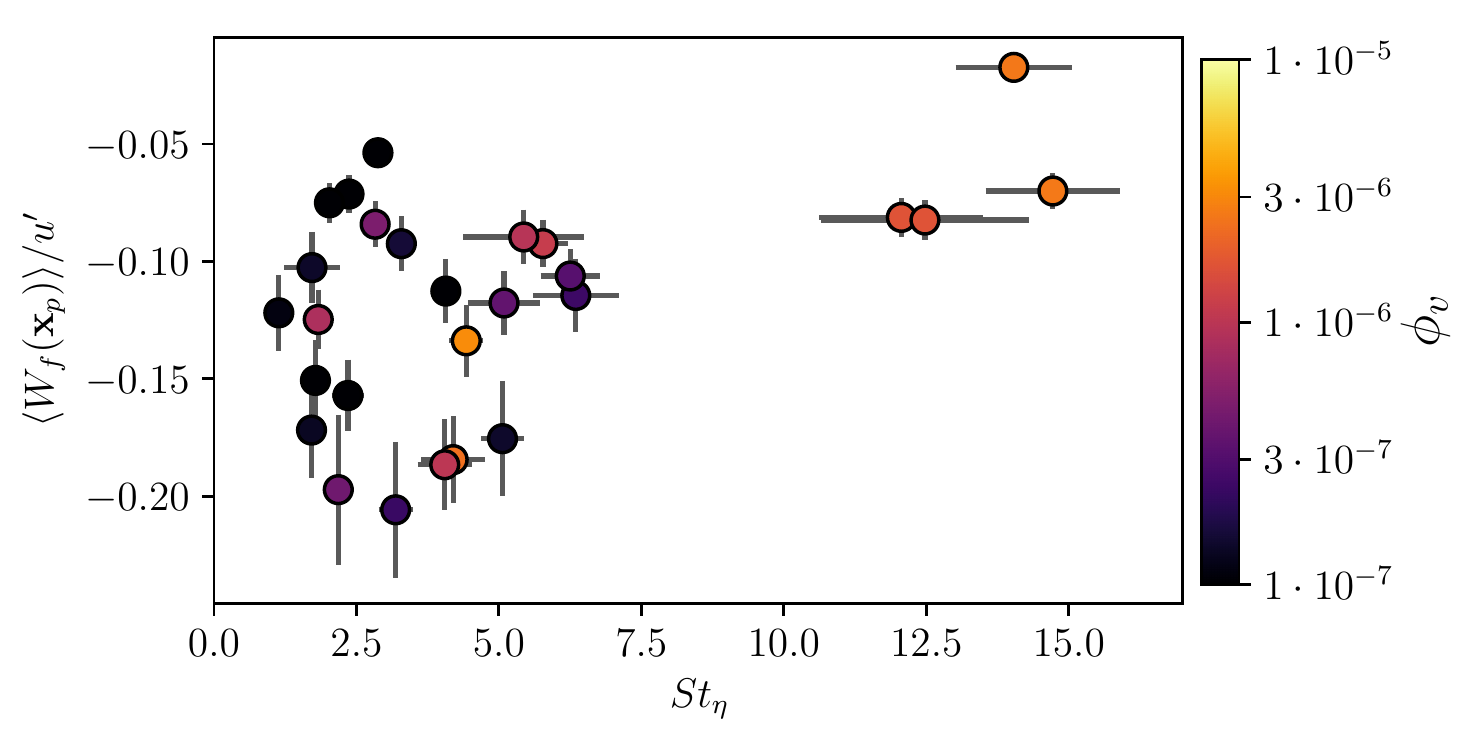}}
		\caption{Vertical fluid velocity at particle location, normalized by the fluid rms velocity, plotted versus particle Stokes number.}
		\label{fig:interp}
	\end{figure}
\end{center}
\indent \indent Comparing figure \ref{fig:settling1}b and \ref{fig:interp}, the oversampling of downward fluid velocity regions seems to account for a large part of the settling enhancement. A more quantitative account can be given in the framework of the point-particle approximation. Retaining only drag and gravity in the particle equation of motion, the fallspeed can be approximated as (Wang \& Maxey 1993):
\begin{equation}
\langle W_s \rangle \approx \langle \mathbf{u_f}(\mathbf{x_p},t)\rangle + \frac{\tau_p g}{\langle f \rangle}
\end{equation}
where $\mathbf{u_f}(\mathbf{x_p})$ is the fluid velocity vector at the particle location obtained via the piecewise-linear interpolation, and $\langle f\rangle$ is the ensemble-average of Schiller \& Naumann correction factor in eq. (2.1), $f = 1 + 0.15Re_p^{0.687}$. We can directly verify this approximation using the instantaneous $Re_p$ measured from the simultaneous PIV/PTV measurements (which will be discussed further in the next section). Figure \ref{fig:settling_comp} shows the ratio between the fallspeed calculated from (5.1) and the measured values. This formulation consistently underpredicts the measured fallspeed. Such a discrepancy between experiments and theory suggests that the one-way coupled point-particle approach, while providing the correct qualitative trend, is missing significant aspects of the particle-fluid interaction. We investigate possible sources of the mismatch in \textsection 5.4, where we consider the instantaneous slip velocity.
\begin{center}
	\begin{figure}
		\centerline{\includegraphics[width=3.5in]{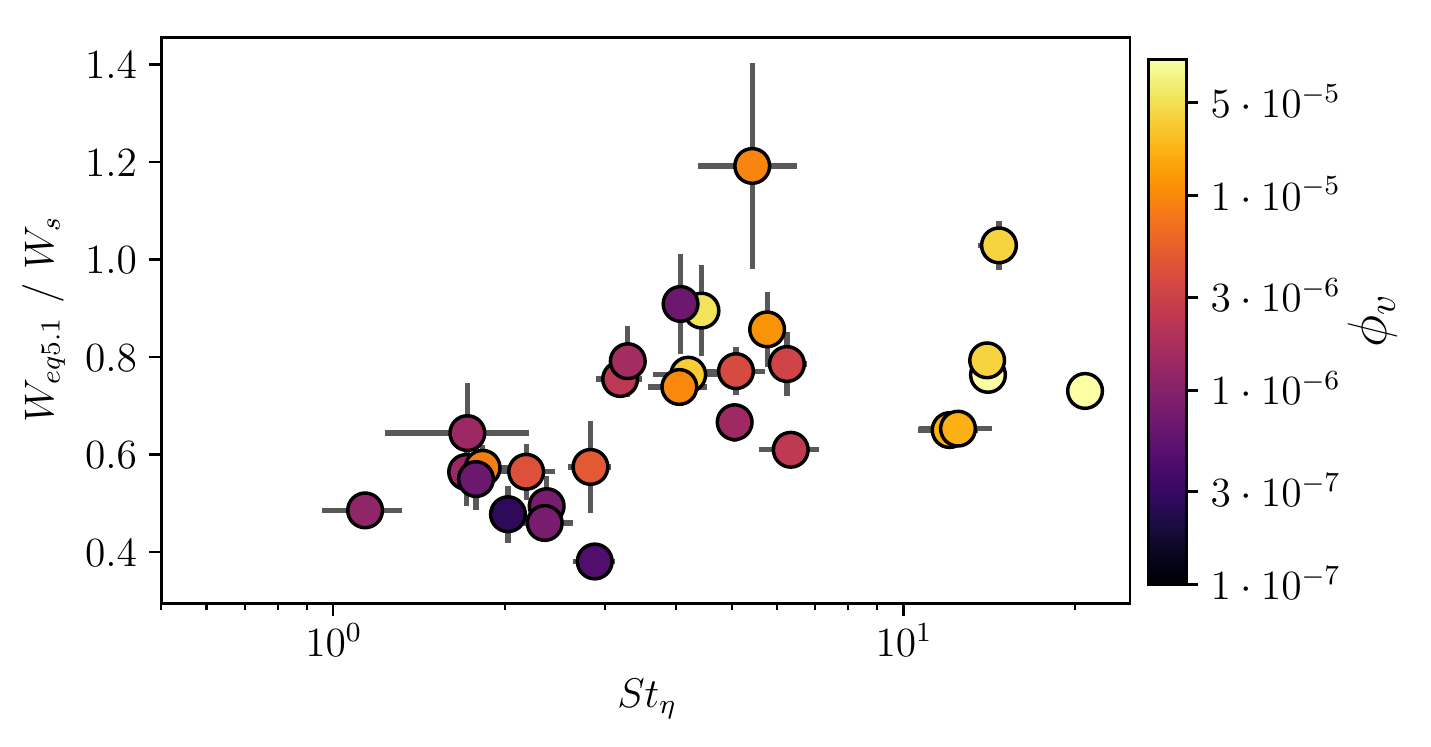}}
		\caption{Prediction of settling velocity based on equation 5.1 normalized by the measured settling velocities, versus particle Stokes number.}
		\label{fig:settling_comp}
	\end{figure}
\end{center}
\subsection{Crossing-trajectory and continuity effect}
\begin{center}
	\begin{figure}
		\centerline{\includegraphics[width=3.5in]{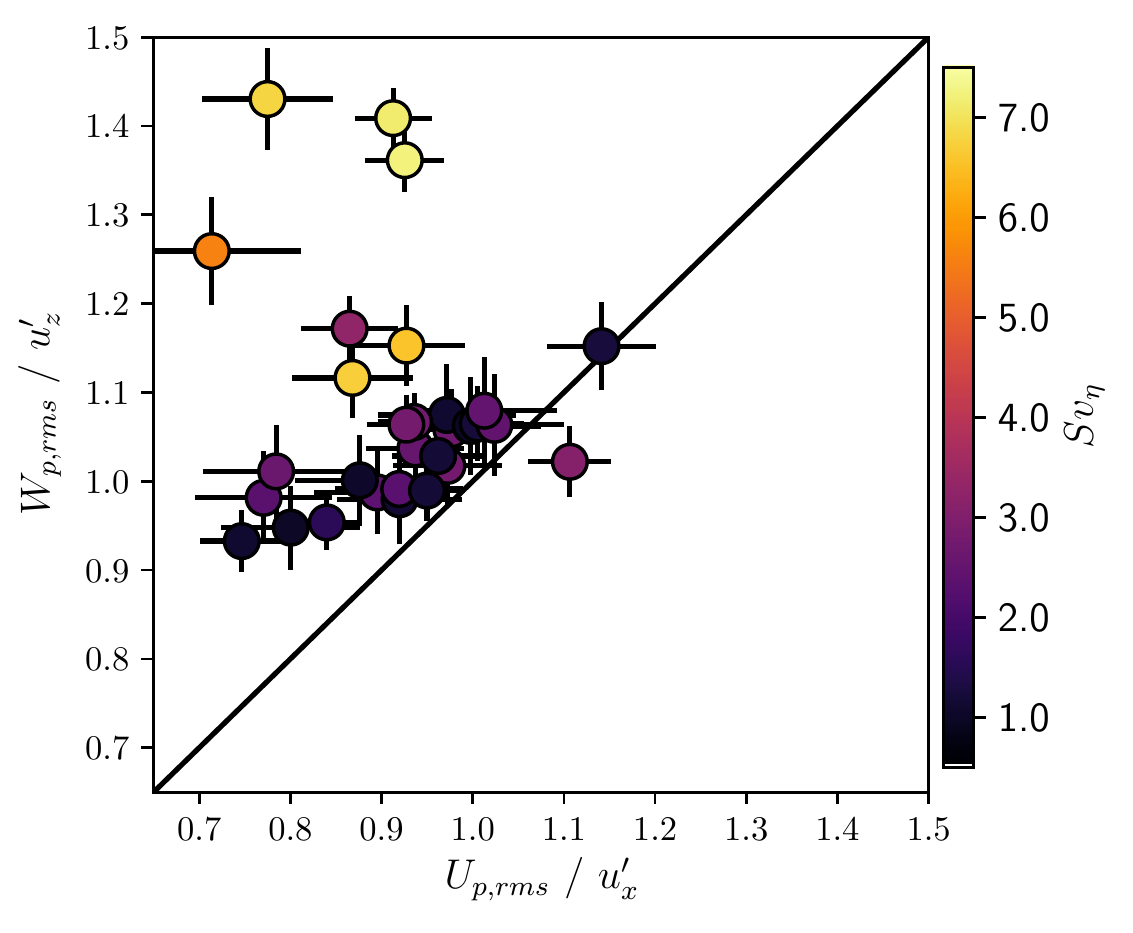}}
		\caption{Ratio of vertical particle to fluid rms velocity versus ratio of the same horizontal velocities.}
		\label{fig:cross}
	\end{figure}
\end{center}

Since \citet{Yudine59}, it has been recognized that the drift induced by body forces such as gravity causes heavy particles to decorrelate from their past velocity faster than a fluid element. This so-called crossing trajectory effect can be quantified by the Lagrangian autocorrelation of the particle velocity \citep{ElghobashiTruesdell92}. For large drift velocities, this reduces to the fluid space-time correlation in an Eulerian frame \citep{Csanady63, SquiresEaton1991b}. In this limit, as the longitudinal integral scale is twice the transverse one, the particle dispersion parallel to the drift direction is double the dispersion perpendicular to it (the so-called continuity effect; \citet{Csanady63, WangStock93}). The footprints of these effects are visible in the Eulerian particle velocities. In figure \ref{fig:cross} we present a scatter plot of the vertical and horizontal r.m.s. particle velocity fluctuations ($W_{p,rms}$ and $U_{p,rms}$, respectively), normalized by the r.m.s. fluid fluctuations in the respective directions ($u^{\prime}_z$ and $u^{\prime}_x$) to account for the anisotropy in our facility. The normalized vertical fluctuations of the particles exceed those in the horizontal direction, the disparity being more substantial for larger $Sv_{\eta}$. This trend is consistent with the continuity effect, and in line with previous analysis of \cite{WangStock93} and measurements of \cite{Good2014}: the falling particles have more time to respond to the vertical fluid fluctuations, due to the larger longitudinal integral scale compared to the transverse one. 
\subsection{Particle-fluid relative velocity}
\begin{center}
	\begin{figure}
		\centerline{\includegraphics[width=\textwidth]{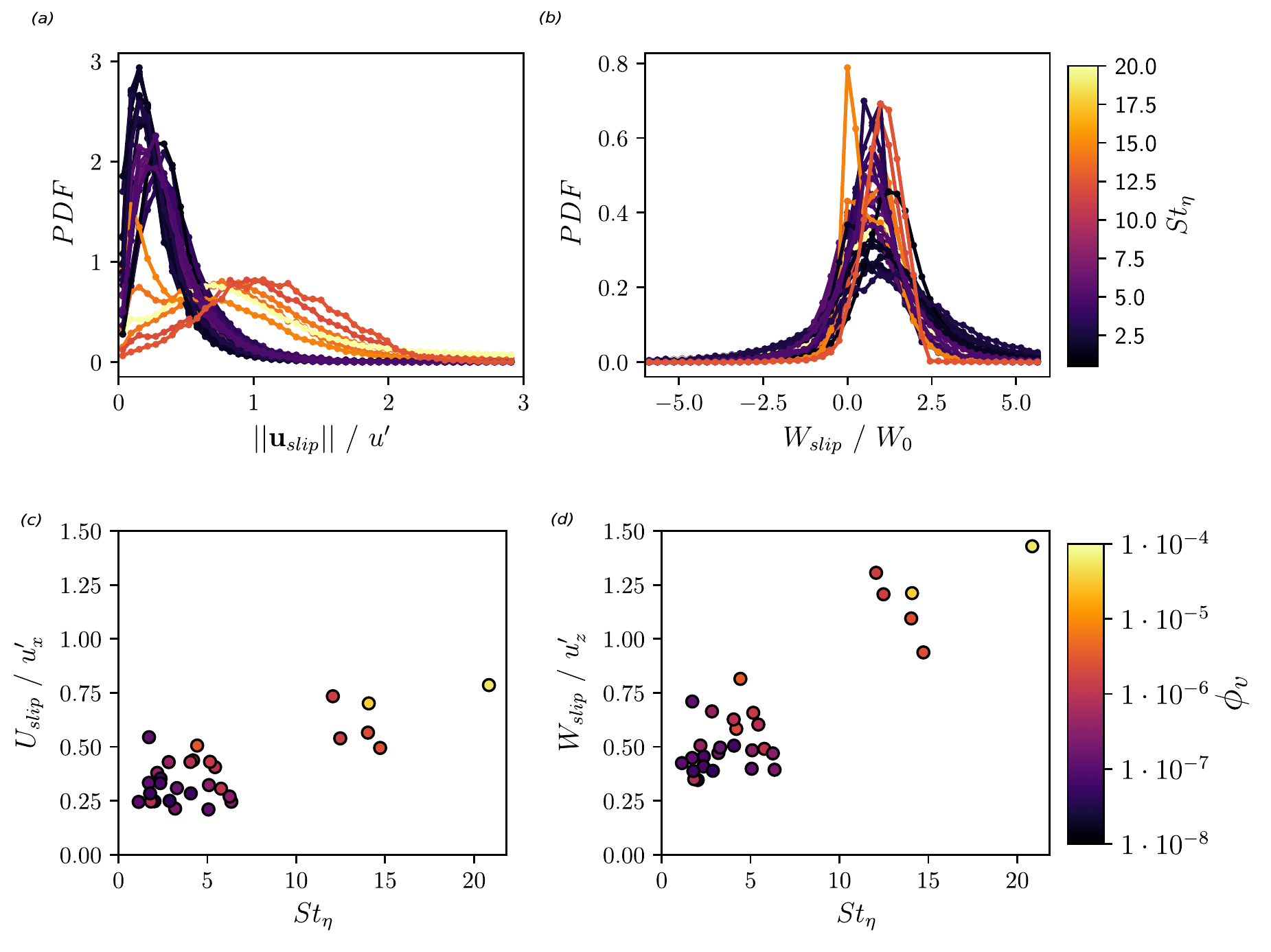}}
		\caption{Probability density functions of slip velocity magnitude normalized by fluid r.m.s. velocity (a), and vertical slip velocity (b) normalized by the still-fluid settling. (c) and (d) plot of horizontal and vertical r.m.s. particle slip velocity, respectively, normalized by the fluid r.m.s. fluctuations along the corresponding direction.}
		\label{fig:slip}
	\end{figure}
\end{center}

\begin{center}
	\begin{figure}
		\centerline{\includegraphics[width=\textwidth]{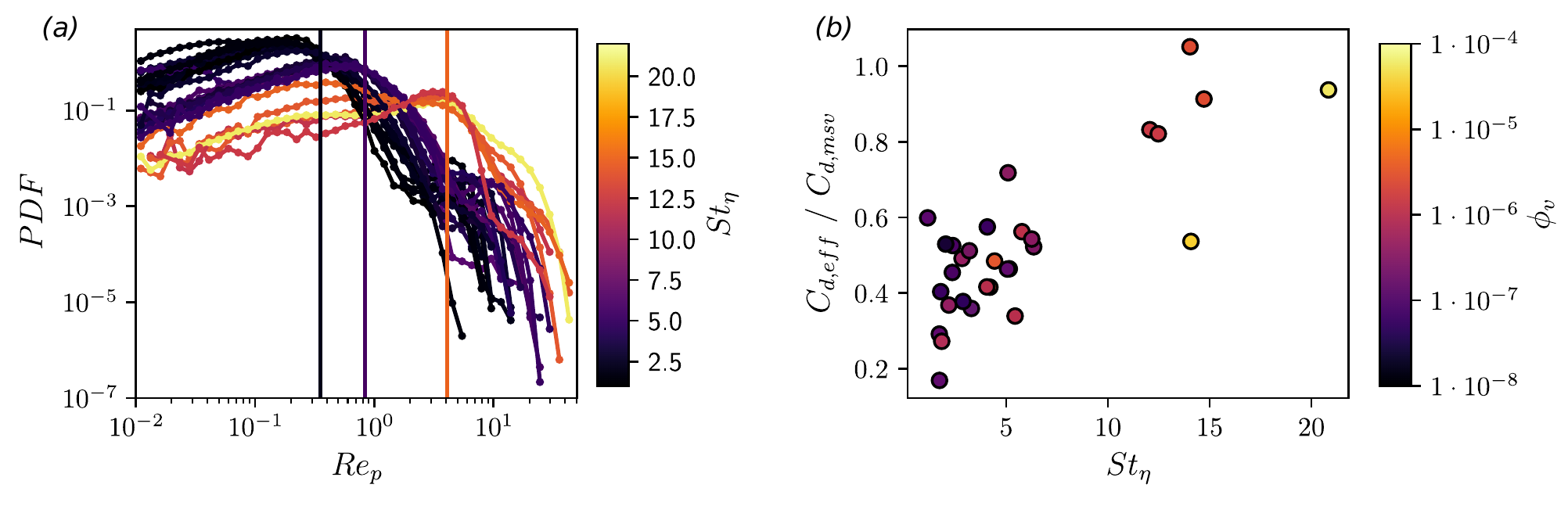}}
		\caption{PDFs of instantaneous particle Reynolds number. Vertical lines represent average Reynolds particle number based on the Stoked still-fluid settling velocity for each respective grouping of experiments. In (b), the ratios between drag coefficients 5.2 and 5.3, showing reduced drag when the "effective drag" model is used.}
		\label{fig:Rep}
	\end{figure}
\end{center}

\begin{center}
	\begin{figure}
		\centerline{\includegraphics[width=\textwidth]{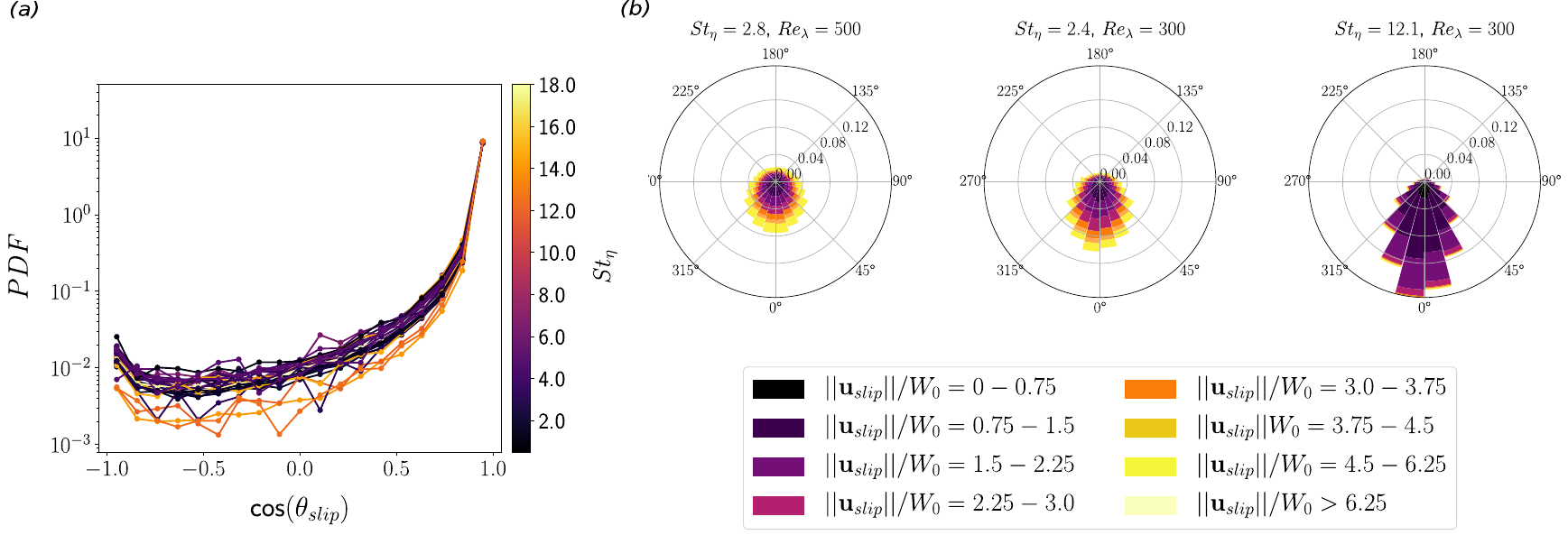}}
		\caption{PDF of the cosine of the angle between the particle velocity and the fluid velocity at the particle location (a). In (b), three example cases showing the probability distribution of the slip velocity orientation and magnitude.}
		\label{fig:slip_or}
	\end{figure}
\end{center}

The relative (slip) velocity between suspended particles and the surrounding fluid plays a vital role in both the particle kinematics and the interphase coupling. The slip velocity determines the exchanged forces and the local flow regime around a particle \citep{MaxeyRiley83, Calzavarini2008, Bellani2012}, and is critical for the momentum two-way coupling with the fluid \citep{Sahu2016}. It is also a key quantity for modeling sub-grid scale dynamics in large-eddy simulations of particle-laden turbulence \citep{Soldati2009}. Due to the abovementioned difficulties in measuring both phases, the slip velocity has been reported by only a few experimental studies, notably \citet{KigerPan2002} and \citet{KhalitovLongmire2003} in wall-bounded turbulent flows, and \citet{YangShy2005} and \citet{Sahu2016} in homogeneous turbulence.\\
\indent Here we evaluate the slip velocity as $\mathbf{u}_{slip} = \mathbf{v_p} - \mathbf{u_f}(\mathbf{x_p})$, where $\mathbf{v_p}$ is the particle velocity vector. In the present measurements we only have access to the in-plane projection of those vectors. Here we only use data from the small and intermediate FOVs. We remark that the relevant definition of slip velocity (as it appears, e.g., in Stokes’ drag law) is the difference between the particle velocity and the undisturbed fluid velocity evaluated at the particle location. Because in the considered cases $d_p < \eta$ and $Re_{p,0} \leq 1$, the region of perturbation from an individual particle (as estimated, e.g., by the Oseen’s solution) is expected to be a few particle diameters, i.e. typically smaller than the Kolmogorov scale. Under this assumption, the interpolated fluid velocity approximates the undisturbed velocity. However, as we shall see, the instantaneous particle Reynolds number $Re_p$ can reach much larger values than $Re_{p,0}$.\\
\indent Figure \ref{fig:slip}a presents PDFs of the slip velocity magnitude normalized by the fluid r.m.s. fluctuation. At the higher $St_{\eta}$, $\mathbf{u}_{slip}$ can significantly exceed $u^{\prime}$, suggesting that the potential energy transferred to the fluid by the fast falling particles is considerable with respect to the turbulent kinetic energy. We will return to this point in \textsection 5.5. In figure \ref{fig:slip}b we plot the PDF of the vertical component of the slip velocity $W_{slip}$, normalized by the still-air particle settling velocity $W_0$. The mean vertical slip is approximately equal to $W_0$, as one expects from (5.1). However, the standard deviation is large and, especially for the lower $St_{\eta}$, greatly exceeds the mean value; indeed, the probability of an upward particle slip is significant. This behavior is consistent with the simulations of particle-laden wall-bounded turbulence by \citet{Zhao2012}, who found that the r.m.s. fluctuations of both streamwise and wall-normal slip velocities were several times greater than their mean. The r.m.s. slip velocities are presented in figure \ref{fig:slip}c,d separating horizontal and vertical components and normalizing by the r.m.s. fluid velocity fluctuations in their respective directions. The slip velocity fluctuations increase with $St$, also consistent with the results of \citet{Zhao2012}: the heavier particles increasingly move independently from the fluid, and their r.m.s. slip velocities reach levels comparable to the r.m.s fluid fluctuations. The vertical r.m.s. is consistently higher than the horizontal. This is likely a consequence of the vertical r.m.s. velocity of the particles being larger than the horizontal one, as discussed in the previous section.
\indent \indent The magnitude of the in-plane slip velocity can be used to calculate the instantaneous particle Reynolds number. The PDFs of $Re_p$ for various cases are plotted in figure \ref{fig:Rep}a, with the vertical lines indicating the respective theoretical $Re_{p,0}$. The large slip variance results in long tails of the distributions: while the wake-shedding regime is never achieved, there is a sizeable probability of $Re_p$ being an order of magnitude larger or smaller than the nominal value. The values are in fact somewhat underestimated, because the out-of-plane slip is not accounted for.\\
\indent The slip velocity and Reynolds number can be used to evaluate the drag coefficient on the settling particles. In presence of background turbulence, this is expected to differ from the “standard drag” in a steady uniform flow \citep{Bagchi2003}. The latter can be estimated following Schiller \& Naumann correction for finite Reynolds number:
\begin{equation}
C_{d,SN} = \frac{24}{Re_{p,0}}\left( 1+0.15Re_{p,0}^{0.687}\right)
\end{equation}
As suggested by \cite{BalaEaton2010}, for particles settling in zero-mean flow homogeneous turbulence, one can evaluate an effective drag coefficient:
\begin{equation}
C_{d,eff} = \frac{24}{\langle Re_v \rangle}\frac{1}{\langle W_p\rangle}\left| \langle \mathbf{u}_{slip} \rangle + 0.15\left( \frac{d_p}{\nu} \right) ^{0.687} \langle\mathbf{u}_{slip}^{1.687}\rangle \right|
\end{equation}
where $Re_v = d_p|\mathbf{v_p}|/\nu$. This may differ from the $C_{d,SN}$ owing to three main factors: the mean fluid velocity seen by the particles being non-zero (e.g., due to preferential sweeping); the non-linear relation between drag and slip velocity; and possible two-way coupling. Most previous studies concerned with the influence of ambient turbulence on $C_d$ considered larger $Re_p$ compared to the present case (typically of order $10^2 - 10^3$; e.g., \citet{WuFaeth94, Warnica1995, Bagchi2003}). In these cases, the effect on the mean drag was found to be small. \citet{Bagchi2003}, however, recognized the potential importance of preferential flow sampling for particles falling freely through turbulence. Here we use (5.2) and (5.3) and plot the ratio $C_{d,eff}/C_{d,SN}$  in figure \ref{fig:Rep}b. There is a significant reduction in effective drag over the $St_{\eta}$ range displaying preferential sweeping. This is consistent with the early measurements of \citet{RudoffBachalo88}, who found substantially reduced drag for liquid droplets in turbulent air. We also observe a clear increasing trend with $St_{\eta}$. This is likely a consequence of diminishing preferential sweeping effects, giving way to those of non-linear drag and possibly loitering \citep{Good2014}. We remark, however, that the high-St data points are affected by larger uncertainty due to two factors: non-negligible perturbation of the local fluid velocity by the particles (because of their relatively large $Re_p$), and limited spatial resolution (because those are imaged with the intermediate FOV). Thus, further analysis and measurements are warranted for those regimes.
\indent\indent We then consider the spatial orientation of the particle slip velocity. Figure \ref{fig:slip_or}a shows the PDF of cos($\theta$), $\theta$ being the angle between $\mathbf{v_p}$ and $\mathbf{u_f}(\mathbf{x_p})$. For moderate particle inertia ($St_{\eta} < 10$) the fluid and particle velocity vectors tend to be closely aligned, with $\theta < 30\deg$ in more than 90\% of the instances. As expected, the more inertial particles are more likely to display a velocity orientation substantially different from the fluid velocity. The probability of oppositely aligned vectors is also non-negligible, possibly due to weak upward gusts that are unable to reverse the particle settling motion. The probability distribution of slip orientation and magnitude is illustrated by the wind rose diagrams in figure \ref{fig:slip_or}b for selected cases. The length of each spoke represents the probability associated to its orientation; additionally, each spoke is broken down in segments representing the probability of a certain slip velocity magnitude (normalized by $W_0$). In all cases the slip velocity favors the downward direction, as expected. In the larger $Re_{\lambda}$ cases, however, the strong fluid velocity fluctuations mitigate this tendency and the azimuthal distribution of the particle slip is less skewed. As discussed above, particles with $St_{\eta} \approx 1$ can have slip velocities several times larger than their still-air fall speed. On the other hand, $\mathbf{u_{slip}}$ for the more inertial cases is closer in magnitude to $W_0$ and mostly oriented downward (typically within $\pm 45\deg$ from the vertical), confirming that these particles follow more ballistic downward trajectories.
\subsection{Turbulence modification by particles}
\begin{center}
	\begin{figure}
		\centerline{\includegraphics[width=\textwidth]{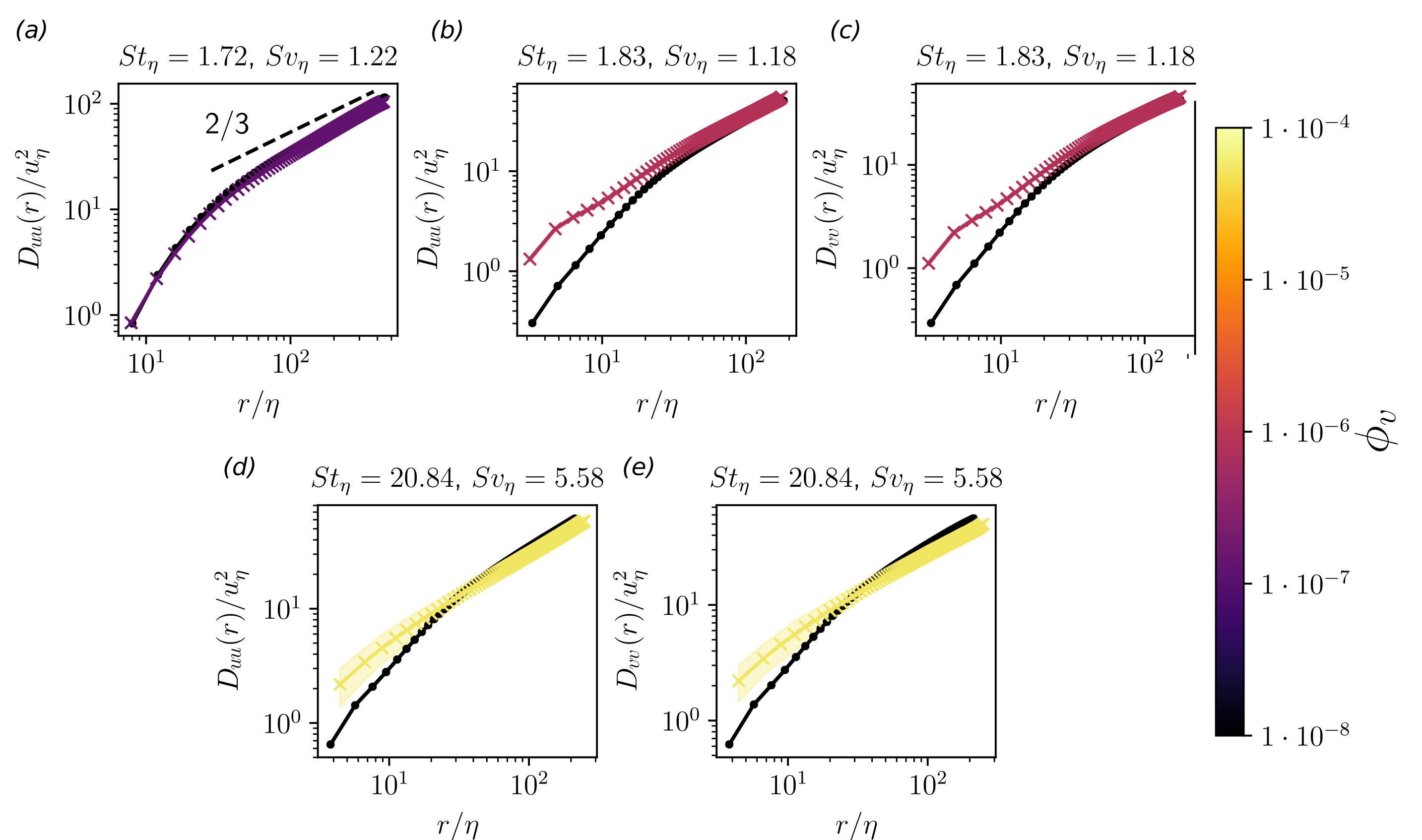}}
		\caption{Horizontal (a,b,d) and vertical (c,e) longitudinal structure functions for laden (color-coded by volume fraction) and unladen (black) multiphase flows. The shaded regions represent 95 \% bootstrap confidence bounds.}
		\label{fig:struct}
	\end{figure}
\end{center}

We finally consider the scale-by-scale effect of the particles on the fluid turbulent fluctuations. We consider the second-order structure functions, which contain analogous information as the energy spectra used in most previous studies:
\begin{equation}
D_{ii}(\mathbf{r})= \overline{[u_i(\mathbf{x} + \mathbf{r}) - u_i(\mathbf{x})]^2}
\end{equation}
Here $\mathbf{x}$ is the generic location on the imaging plane, $\mathbf{r}$ is the separation vector, and $u_i$ is the $i$-th component of the velocity fluctuations. Thus, we denote as $D_{uu}$ and $D_{ww}$ the structure functions associated to the horizontal and vertical components. We focus on longitudinal structure functions (i.e., with separations parallel to the velocity components), the transverse structure functions yielding similar results. For this analysis we use only small and intermediate FOVs, in order to resolve a sufficient fraction of the fine-scale fluctuations.\\
\indent Figure \ref{fig:struct} displays horizontal and vertical structure functions for representative cases, comparing the laden and unladen measurements. At very low volume fractions ($\phi_v = O(10^{-7})$), the particles marginally affect the energy distribution across scales, and the expected scaling $D_{ii}(r) \sim r^{2/3}$ is approached entering the inertial range according to Kolmogorov theory (Fig. \ref{fig:struct}a). At significant loadings ($\phi_v = O(10^{-5})$), we observe an increase of turbulent energy at small scales. For $St_{\eta} = O(1)$ (figure \ref{fig:struct}b)  the laden and unladen curves approach each other at larger scales, possibly leading to a cross-over which however is not captured within the field of view. At $St_{\eta} = O(10)$, the cross-over happens at smaller scales as demonstrated in figure \ref{fig:struct}d, in which particles appear to excite turbulent fluctuations at the small scales and modulate them at the large scales. This behavior (sometimes termed “pivoting”) was reported by several numerical studies, as reviewed in detail by \citet{PoelmaOoms2006}. Most of those \citep[e.g.,][]{SundaramCollins1999, Ferrante2003} showed the cross-over wavenumber to increase with increasing particle response time, consistent with our observations. The only previous experimental study to clearly demonstrate the pivoting effect was, to our best knowledge, the one from \citet{PoelmaOomsWesterweel2007}. However, a direct comparison is difficult as these authors considered spatially decaying turbulence at maximum $Re_{\lambda} \approx 29$.
\indent\indent The vertical velocity structure functions (figures \ref{fig:struct}c,e) display the same behavior as their horizontal counterpart (figures \ref{fig:struct}b,d), suggesting that the turbulence responds to the presence of the particles in similar ways in both directions. It is indeed verified that the anisotropy ratio $u_x^{\prime}/u_z^{\prime}$ is not significantly affected by the particles. \citet{PoelmaOomsWesterweel2007} and \citet{Frankel2016} found that particles enhanced the fluid velocity fluctuations aligned with gravity compared to the transverse ones, whereas in \cite{HwangEaton2006b} the level of isotropy was marginally influenced even at particle mass loadings that heavily modulated the turbulence intensity. Among these studies, Hwang \& Eaton considered a much higher Reynolds number ($Re_{\lambda} \approx 240$), comparable to the present study. One may conjecture that the wide range of scales allows for a redistribution of the energy among the different components; certainly, further studies that systematically investigate the role of $Re_{\lambda}$ are needed.\\
\indent\indent Even for the higher volume fractions, the changes in the fluid turbulence intensities in both vertical and horizontal directions (not presented) are scattered within the measurement uncertainty, with no consistent trend in terms of $St_{\eta}$, $Sv_{\eta}$, or $\phi_v$. This suggests that the turbulence augmentation/attenuation, over different scales and from different mechanisms, largely compensate each other, at least in the considered range of parameters. On one hand, particles increase the overall inertia of the mixture and locally enhance the dissipation around them, modulating the turbulence; on the other hand, the drag force exerted by the falling particles perturbs the fluid an excites the turbulence \citep{BalaEaton2010}. Both effects are amplified for larger and heavier particles (which provide higher loading, distort more the flow around them, and shed more energetic wakes). For comparison, \citet{HwangEaton2006} investigated particles with $St_{\eta} \approx 50$ settling in homogeneous turbulence and measured a reduction of r.m.s. fluid velocity of about $10-15\%$ for a mass fraction of 0.1 (close to the highest loading considered in the present measurements). The balance between opposing effects was demonstrated by measurements obtained in the same facility in micro-gravity \citep{HwangEaton2006b}, where the turbulence modulation was found to be greater than in the fixed laboratory frame. \citet{HwangEaton2006b} also compared their laboratory results against DNS studies which showed significantly smaller turbulence attenuation. While this suggested that the point-particle approach used in the simulations missed important physics, the DNS were also at significantly smaller $St_{\eta}$, between 1 and 11 and comparable to the present cases.
\section{Discussion and conclusions}
The present experimental approach has allowed us to gain insight into several outstanding issues in particle-laden turbulence. The jet-stirred homogeneous air turbulence chamber is particularly suitable to characterize the considered regime: the lack of mean flow enables the unbiased measurement of the settling velocity, also yielding a much larger dynamic range for the velocity fluctuations compared to wind tunnel experiments \citep{CarterColetti2017}. Moreover, the large region of homogeneous turbulence is crucial for the particles to interact with the full range of turbulent scales \citep{BellaniVariano2014}. The simultaneous imaging of both phases allows the characterization of their interplay, within the inherent limits posed by the imaging accuracy and its planar nature. Finally, the relatively high $Re_{\lambda}$ warrants the separation of scales needed to identify the dominant flow parameters.\\
\indent We have characterized the particle spatial distribution using both RDF and Vorono\"i diagrams. The former show a power-law decay over (and beyond) the near-dissipative scales, followed by a long exponential tail. The latter indicates that clustering extends far beyond the dissipation range, although the particle field does not display scale-invariant properties over such distances. It is confirmed that clustering is most intense for $St_{\eta} \approx 1$, but remains significant even for more inertial particles, which in fact cluster over larger regions. This is due partly to their response time being comparable to the time scales of larger eddies, and partly to the effect of gravity. It was indeed reported in several numerical investigations that gravitational settling enhances clustering for $St_{\eta} > 1$ \citep[see][]{IrelandBraggCollins2016b}. Gravity is unavoidable in laboratory experiments, thus its effect is hard to discern from inertia. However, our results for various $St_{\eta}$ and $Sv_{\eta}$ are at least consistent with this trend. We argue that such an effect is associated to $u^{\prime}$ becoming the relevant velocity scale for the interaction of turbulence with fast-falling particles: the latter quickly decorrelate from Kolmogorov-size eddies, and rather respond to large-scale velocity fluctuations. Indeed, the fraction of clustered particles and the cluster size generally increases with particle inertia.\\
\indent The Vorono\"i diagram method generally confirms the findings from the RDF analysis, and allows us to explore further aspects of the particle distribution, focusing on individual clusters. The properties of those discrete groups of highly concentrated particles bears particular significance, as these may interact with one other and, by virtue of their collective action, modify the surrounding flow \citep{Monchaux2012}. We find that sufficiently large clusters (i.e. “coherent clusters” in the definition of \citet{Baker2017} follow power-law size distributions over several decades, and display a precise fractal dimension. Therefore, borrowing the terminology proposed by \citet{Paola2009} to describe scale-invariance in geomorphology, these objects exhibit both internal similarity (between the system and small parts of it) and external similarity (between the system and small copies of it). This also suggests that they follow (without necessarily replicating) the self-similar topology of the underlying turbulence. The clusters are usually elongated and often aligned with the vertical direction. The mean particle concentration in a cluster can be an order of magnitude higher than the average, and is largely independent on the cluster size.\\
\indent It may appear surprising that particle clusters display both internal and external similarity, and over such a wide range of scales. After all, turbulence is known to display strong departures from self-similarity associated to its intermittent behavior. However, a few considerations are in order. First, intermittency is most evident in high-order quantities in which the size and shape of the clusters of inertial particles might be relatively insensitive. In fact, the intense vortex filaments usually regarded as responsible for scale-variant behaviors are precisely the structures where particles of  $St_{\eta} = O(1)$ are unlikely to be found. Also, by focusing on coherent clusters, we set a size threshold larger than the scale over which intermittency is manifest.  One point that deserves deeper investigation is that, unlike the Vorono\"{i}-based cluster topology, two-point measures of clustering such as RDFs do not display scale-invariance in the inertial range (as also confirmed by our experiments). \citet{Bragg2015} analyzed the scaling of RDFs in this range, and their arguments may also be the basis of a theoretical investigation of the inertial-scale cluster dynamics; this however is outside the scope of the present work.\\
\indent One of the major findings of the study is that, for $St_{\eta}$ and $Sv_{\eta}$ of order unity and in the considered range of $Re_{\lambda}$, turbulence produces almost a three-fold increase in settling velocity with respect to quiescent conditions. Such a dramatic effect has far-reaching consequences in a myriad of natural and engineering settings, from atmospheric precipitation to industrial processes and pollutant dispersion. Remarkably, the effects of turbulence on the fallspeed of hydrometeors have only recently begun to be addressed in field studies \citep{Garrett2014}. In particular, the present results are consistent with the measurements by \citet{Nemes2017} who imaged snowflakes settling in the atmospheric surface layer. The data is best collapsed by a mixed scaling based on $\tau_{\eta}$ and $u^{\prime}$, which may reflect how the latter is indeed the relevant velocity scale ruling the settling enhancement. This argument is consistent with the idea of distant-scale interaction, which has been used in wall-bounded flows to explain the success of mixed scaling of the streamwise turbulent fluctuations \citep{Marusic2003}. \\
\indent Clusters fall even faster than the average particle in turbulence. In the considered range of parameters, this appears to be due to the particle tendency to oversample downward flow regions, according to the preferential sweeping mechanism. The latter is shown to be responsible for the observed enhanced settling, and in the present regime is a more recognizable feature than the preferential concentration in high-strain/low-vorticity regions. Owing to the crossing trajectory and continuity effects, the vertical velocity fluctuations of the particles are stronger than the horizontal ones. The r.m.s. of the vertical slip velocities are also larger than the horizontal, and both can largely exceed their respective mean values. Thus, the actual particle Reynolds number can greatly differ from the nominal value usually used to correct Stokes’ law. Indeed, the effective drag coefficient calculated via the simultaneous two-phase measurements is found to depart significantly from the standard estimate. This points to clear limitations of the point-particle approach, perhaps accounting for some of the quantitative disagreement between experiments and DNS at matching conditions \citep{Good2014}.\\
\indent At sufficient concentrations, the particles excite small-scale fluid velocity fluctuations, which however may represent a small fraction of the turbulent kinetic energy. For the heavier particles, we generally observe a reduction of turbulent energy at the larger scales. Over the considered range of parameters, the measurements did not indicate a trend of augmentation versus attenuation of the overall turbulence intensity. This is perhaps not surprising, considering the multiple counteracting effects contributing to turbulence modification at the different scales, along with the difficulty of varying one parameter at a time. Dedicated experiments in which the volume fraction is systematically varied over a wide range, keeping all other settings constant, stand a higher chance of providing some conclusive answer. This was not the main goal of the present work, and it will be the focus of a future campaign. Such a study, however, is expected to be challenging: for sub-Kolmogorov particles and relatively high $Re_{\lambda}$, substantial mass loadings require large number densities, posing problems for the fluid velocimetry. Those can be alleviated by imaging at very high resolution; but this limits the FOV, and consequently the insight into multi-scale mechanisms. Therefore, the use of very large sensors and/or the simultaneous deployment of multiple cameras appear necessary to fully quantify two-way coupling effects in preferentially concentrated particle-laden turbulence. \\
\indent Overcoming the limitations of planar measurements will require a leap in performance of 3D particle imaging at high concentrations. Novel approaches based on Tomographic PTV are breaking grounds in single-phase velocimetry \citep{Schanz2016}, and may be applied to inertial particles. Advanced numerical methods are also expected to shed new light: particle-resolved DNS has been successfully applied to investigate relatively large particles \citep{Naso2010, Lucci2010, Tenneti2014, Cisse2013, GaoLiWang2013, Uhlmann2014, Fornari2016}, and recently even Kolmogorov-sized particles (Schneiders, Meinke \& Schroeder 2017). The ever-increasing computational power will soon allow resolving even smaller objects, elucidating the role of mechanisms such as the local hydrodynamic interaction between particles \citep{WangAyalaGrabowski2007}. Still, capturing truly collective effects with this approach will require enormous computational resources. Considering the present case as an example (e.g., $L \approx 100$ mm and $d_p \approx 50\mu$m), a cubic domain of size $2\pi L$ with a volume fraction of $5\cdot 10^{-5}$ contains $O(10^8)$ particles. Low-order methods able to incorporate the essential physics are therefore sorely needed; the role of experiments and particle-resolved simulations in informing and validating them will be essential.\\

\textbf{Acknowledgements}\\
\indent The present work was supported in part by the U.S. Army Research Office (Division of Earth Materials and Processes, grant W911NF-17-1-0366). AJP is supported by a National Science Foundation Graduate Research Fellowship. LJB is supported by a National Defense Science and Engineering Graduate Fellowship. The manuscript benefited from the insightful comments of the anonymous referees. We are thankful to Douglas Carter and Omid Amili for valuable help during the measurement campaign, and many fruitful discussions. Alec would especially like to thank Andras Nemes for all his guidance and help on the software side of things. Lastly, a small thanks to Jason Herzfeld and Gary Petersen, both who helped run an experiment or two when they were supposed to be having fun visiting Alec.  
\bibliographystyle{jfm}

\bibliography{submission_draft}

\end{document}